\documentclass[acmtog,nonacm,balance=false]{acmart}

\usepackage{booktabs} 

\usepackage{algorithm}      
\usepackage{algpseudocode}  
\floatname{algorithm}{ALGORITHM}

\PassOptionsToPackage{table,dvipsnames}{xcolor}
\usepackage{pifont}
\usepackage{caption}
\usepackage{enumitem}
\usepackage{subcaption}
\usepackage{outlines}
\usepackage{cuted}
\usepackage{hyperref}
\usepackage{xcolor}
\usepackage{colortbl}
\usepackage{bm}
\usepackage{amsmath}
\usepackage{array}
\usepackage{multirow}
\usepackage{tabularx}
\usepackage{todonotes}
\usepackage{dsfont} 

\newcommand\mypara[1]{\vspace{4pt}\noindent\textbf{#1.}}

\usepackage[capitalize]{cleveref}
\crefname{section}{Sec.}{Secs.}
\Crefname{section}{Section}{Sections}
\Crefname{table}{Table}{Tables}
\crefname{table}{Tab.}{Tabs.}
\definecolor{Cerulean}{rgb}{0.0, 0.48, 0.65}
\definecolor{myred}{rgb}{1, 0.6, 0.6}
\definecolor{myyellow}{rgb}{1,1, 0.6}
\definecolor{myorange}{rgb}{1, 0.8, 0.6}
\definecolor{mycolor_blue}{HTML}{E7EFFA}
\definecolor{mycolor_green}{HTML}{E6F8E0}
\definecolor{mycolor_gray}{HTML}{ECECEC}
\definecolor{pearDark}{HTML}{2980B9}

\acmJournal{TOG}

\begin{document}
\title{Hi-TOPS: Hierarchical Topology-aware Scoring Prior for 3D Part Decomposition}


\author{Ruoyu Wu}
\authornote{Equal Contribution}
\email{ruoyu.wu2@unsw.edu.au}
\affiliation{%
  \institution{University of New South Wales}
  \city{Canberra}
  \country{Australia}
}

\author{Zhenhong Sun}
\authornotemark[1]
\email{zhenhong.sun@anu.edu.au}
\affiliation{%
  \institution{Australian National University}
  \city{Canberra}
  \country{Australia}
}

\author{Xiaoming Gong}
\email{522025150053@smail.nju.edu.cn}
\affiliation{%
  \institution{Nanjing University}
  \city{Nanjing}
  \country{China}
}

\author{Yuxin Xian}
\email{yuxin_xian@163.com}
\affiliation{%
  \institution{Southwestern University of Finance and Economics}
  \city{Chengdu}
  \country{China}
}

\author{Zhi Wang}
\email{zhiwang@nju.edu.cn}
\affiliation{%
  \institution{Nanjing University}
  \city{Nanjing}
  \country{China}
}

\author{Yawen Chen}
\email{wendy.chen1@unsw.edu.au}
\affiliation{%
  \institution{University of New South Wales}
  \city{Canberra}
  \country{Australia}
}

\author{Huadong Mo}
\authornote{Corresponding author.}
\email{huadong.mo@unsw.edu.au}
\affiliation{%
  \institution{University of New South Wales}
  \city{Canberra}
  \country{Australia}
}

\author{Daoyi Dong}
\email{daoyidong@gmail.com}
\affiliation{%
  \institution{University of Technology Sydney}
  \city{Sydney}
  \country{Australia}
}

\renewcommand{\shortauthors}{Wu et al.}

\begin{abstract}
Accurate 3D part decomposition requires separating shapes into structurally meaningful components with precise boundaries while preserving articulation seams and thin attachments. However, existing approaches often suffer from a \emph{structural-scale mismatch}: the geometric evidence for part separation is most reliable at the meso scale, yet many pipelines operate either too globally to respect joints or too locally to remain robust to noise, leading to joint bridging and thin-structure omission. 
We propose \textbf{Hi-TOPS}, a \textbf{Hi}erarchical \textbf{Top}ology-aware \textbf{S}coring Prior where topology-aware denotes structure-level connectivity and boundary-relevant geometric organization rather than formal invariants from algebraic topology or topological data analysis. It discretizes the normalized shape into meso-blocks and aggregates complementary intrinsic cues into a \textit{Flow-Freeze} field.
Flow areas define expandable support volumes that guide primitive coverage, while freeze areas mark articulation-sensitive and thin-structure boundaries that restrict primitive growth.
To improve meso-scale coverage across varying local complexity, we compute this field at multiple voxel resolutions and perform rule-based freeze fusion, forming a hierarchical Flow-Freeze prior that refines freeze boundaries.
Building on this prior, we introduce a \textbf{TSDF-guided body-surface SuperQuadric (SQ) fitting} that seeds primitives from the farthest interior anchors to capture dominant cores while simultaneously recovering residual surface structures that are underrepresented by distance-based maxima.
Finally, an SQ-to-mesh assignment transfers voxel-space primitives back to connected surface components, preserving freeze-dominant boundary patches as residual details.
Experiments across diverse shapes demonstrate that Hi-TOPS produces stable decompositions with editable primitive parts and high-fidelity residual patches, enabling reliable abstraction for downstream editing and control. Code is available at \url{https://engineeringai-lab.github.io/HITOPS/}.

\end{abstract}


%
%


%
%


\begin{teaserfigure}
    \centering
    \includegraphics[width=0.86\textwidth]{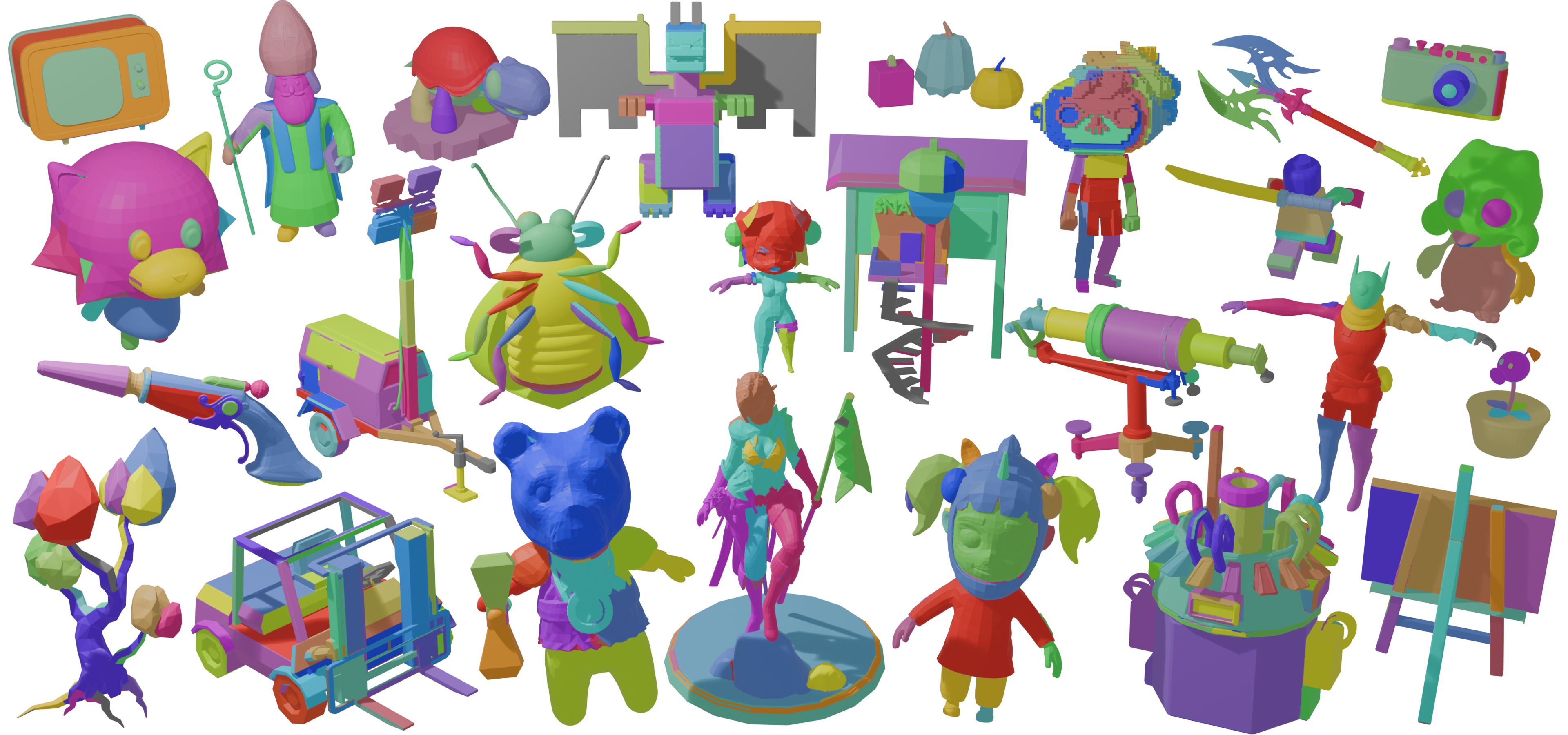}
    \caption{We showcase overall segmentation results in a composite scene, demonstrating the robustness of our \textbf{Hi-TOPS}.}
    \label{fig:teaser}
    \vspace{5pt}
\end{teaserfigure}

\maketitle

\section{Introduction}
\label{sec:intro}

3D part segmentation aims to decompose unconstrained meshes into structurally meaningful components defined by \emph{precise and stable boundaries} \cite{sampart3d_yang2024, p3samma2025,  partsam_zhu2025}. These decompositions serve as the geometric foundation for articulated manipulation and controllable generation~\cite{xpart_2024, holopart_yang2025, omnipart_2025yang}. In this work, topology denotes the part-level structural organization of a shape—how volumetrically coherent regions are connected and where narrow joints, thin attachments, or local constrictions provide geometric evidence for separation—rather than formal topological invariants. Hi-TOPS uses a geometry-driven score as a proxy for such boundary-relevant structure. Crucially, the utility of a segmentation is determined by its structural fidelity: algorithms must successfully disentangle narrow joints and thin attachments while preserving the integrity of the object's main body \cite{reachFeature_2014, sgpn_2018, partnet_2019}. This requirement creates a paradox: the segmentation must be semantically coherent (understanding the whole) yet geometrically sharp (respecting local transitions), but existing methods often sacrifice one objective for the other.

The failure of current approaches to balance these objectives can be traced to a recurring issue: the improper choice of \emph{operating scale}. Top-down methods, including foundation model-based approaches \cite{sam_2023, dinov2_oquab2023} and volumetric primitive fitting \cite{mps_2023, lightSQ, learningShapeAbstraction_2017, robustShapeFittingfor3dSceneAbs_2024}, prioritize global coherence. Consequently, their inference is often too coarse or view-dependent to detect the subtle "stopping signals" at articulations, causing parts to bleed across joints. Conversely, bottom-up geometric methods \cite{3dmeshSegMeanShiftedCurv_2008, Yamauchi2005Mesh, vsa_2004, meshpartition_1999mangan} rely on local curvature and normal cues. While these methods capture high-frequency details, they are inherently vulnerable to mosaic noise and lack the structural context to form closed partitions. This dichotomy reveals a fundamental \textbf{structural-scale mismatch}: the critical evidence for articulated segmentation resides at a meso-scale, falling within the blind spot between the global bias of semantics and the local volatility of geometry.

Motivated by this, we revisit 3D decomposition as a \emph{structural prior construction} problem, seeking a geometric representation that can function as both a direct solver and a robust supervisory signal for learning paradigms \cite{wu2025sonata}. We empirically examine block-based partitioning across granularities and observe that aggregating intrinsic cues at an appropriate \emph{meso-scale} yields a stable structural field. This field effectively governs the segmentation dynamics, revealing where decomposition should \emph{flow} to capture volumetric coherence and where it must \emph{freeze} to respect geometric bottlenecks. This indicates that selecting the correct scale is crucial for reliable boundaries, not merely an implementation detail.

Based on these observations, we propose an \textbf{Hi}erarchical \textbf{Top}ology-aware \textbf{S}coring Prior for 3D part decomposition, abbreviated as \textbf{Hi-TOPS}.
It discretizes the normalized shape volume into \textit{meso-scale blocks} and computes a score per block by aggregating complementary intrinsic cues into a robust measure of structural complexity.
The resulting score field induces a multi-resolution partition with explicit semantic roles: low-complexity areas form \textbf{Flow} support areas that facilitate explanations by primitives, while high-complexity areas form \textbf{Freeze} areas that prevent cross-joint intrusion and protect articulation boundaries.
To improve meso-scale coverage across shape complexity, we compute this field at multiple voxel resolutions and perform rule-based Freeze fusion, forming a hierarchical Flow-Freeze prior.
In this way, Hi-TOPS bridges noisy local difference and global part-level abstraction with a topology-aware prior.

Building on the hierarchical prior, we design a \textbf{TSDF-guided body-surface SuperQuadric (SQ) fitting} strategy.
Within the Flow area, we adopt a sequential initialization scheme: we prioritize the farthest interior anchors to first consolidate dominant volumetric cores (the body) and subsequently transition to anchors on the remaining surface to systematically recover thin structures or small components that are underrepresented by interior distance maxima. 
During SQ inflation, each primitive uses TSDF samples from a one-voxel outward shell for parameter updates, while expansions intersecting Freeze areas are rejected to stop growth at boundaries. 
Finally, to achieve accurate surface boundaries in the original mesh domain, we introduce an SQ-to-mesh assignment strategy. This module robustly transfers voxel-space primitive groups back to connected mesh components via geometric association while preserving Freeze-dominant boundary patches for fine detail.

The main contributions are summarized as follows:
\begin{itemize}[leftmargin=*, noitemsep, nolistsep]
    \item[$\bullet$] We analyze the \textbf{structural-scale mismatch} in geometry-driven
    part decomposition.
    \item[$\bullet$] We introduce a hierarchical \textbf{meso-scale} Flow--Freeze prior
    from complementary intrinsic cues.
    \item[$\bullet$] We develop TSDF-guided \textbf{Body-Surface SQ fitting} and
    SQ-to-mesh assignment for editable mesh decompositions.
\end{itemize}
\begin{figure*}[t]
    \centering
    \includegraphics[width=0.96\textwidth]{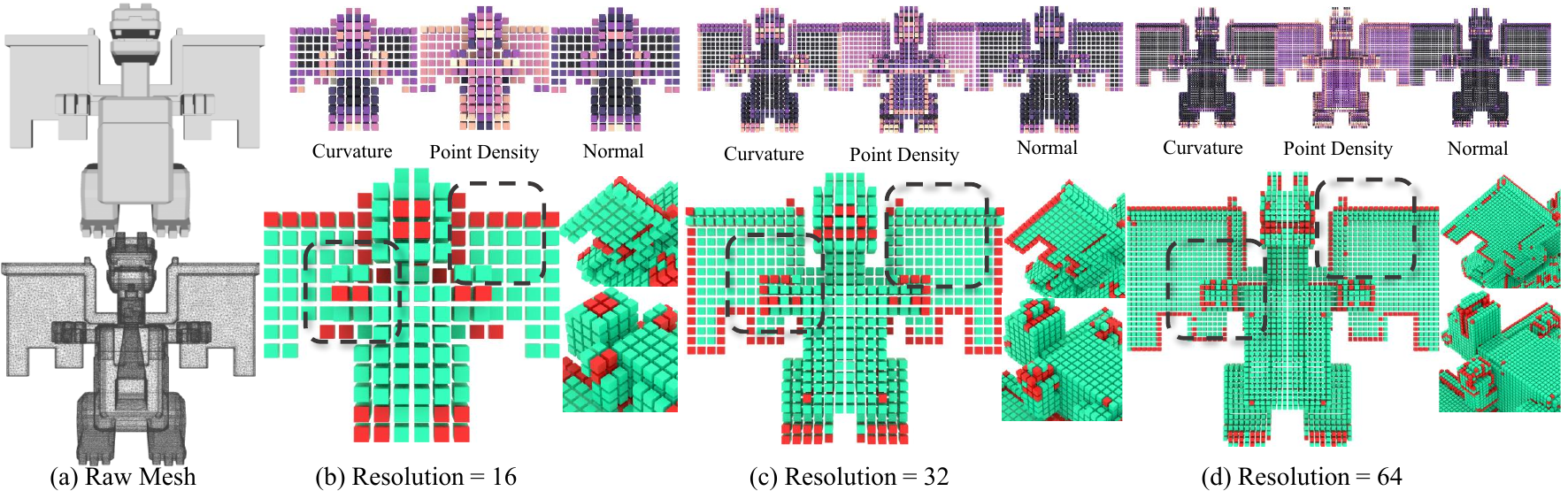}
    \caption{
    \textbf{Multi-scale voxelization and intrinsic cues.}
    The leftmost panels show the input mesh and its wireframe complexity.
    Different voxel resolutions provide complementary evidence: coarse levels ($r=16/32$) capture stable volumetric support, while finer levels ($r=64$) reveal narrow joints and thin boundary details.
    The top panels visualize the three intrinsic geometric cues (Curvature, Normal, and Density), used to construct our Hi-TOPS prior.
    }
    \label{fig:res_comp}
\end{figure*}

\section{Related Work}
\label{sec:related}

\mypara{Learning-based 3D segmentation with 2D priors}
Recent methods either learn directly from native 3D data ~\cite{pointnet2017, p3samma2025, partsam_zhu2025} or transfer 2D foundation-model priors to 3D through multi-view rendering and feature lifting ~\cite{partslip_2023, partslip_pp_2023, sampart3d_yang2024, parfield_2025, mirageroom_2025, pointsam_zhou2025, find3d_2025_ICCV, s2am3d_su2025, cops_garosi2025}. Such priors provide semantic and open-vocabulary cues when labeled 3D data are scarce ~\cite{sam_2023, dinov2_oquab2023, partdistill_2023}. In contrast, Hi-TOPS requires neither training data nor 2D priors, constructing its hierarchical Flow--Freeze prior solely from intrinsic mesh geometry.

\mypara{Primitive-based abstraction and part decomposition}
Primitive-based representations (e.g., cuboids, cylinders, superquadrics, convex parts) provide compact, editable abstractions~\cite{lightSQ, primAny, ems_2022, mps_2023}. Most methods rely on iterative fitting, split-and-merge, or volumetric optimization, but may bridge articulations or miss thin structures without explicit structural stopping constraints~\cite{mps_2023, ems_2022, primAny}. We introduce a hierarchical Flow-Freeze prior with constrained inflation, allowing growth in Flow areas while Freeze areas prevent bridging at joints.

\mypara{Geometry-driven boundary detection and intrinsic cues}
Classical geometry processing approaches aim to identify part boundaries using intrinsic signals such as curvature, normal variation, and other differential or spectral descriptors \cite{3dmeshSegMeanShiftedCurv_2008, Yamauchi2005Mesh, vsa_2004, meshpartition_1999mangan}.
These cues can be computed without training data and often correlate with articulations and functional separations.
However, purely local cues are sensitive to tessellation noise and typically yield fragmented, high-frequency boundary evidence that is difficult to consolidate into coherent part partitions or optimization constraints \cite{3dmeshSegMeanShiftedCurv_2008}.
In contrast, we aggregate complementary intrinsic cues into a robust meso-scale score over macro-blocks, producing a volumetric structural field that is stable and directly actionable for structure-aware 3D part decomposition.
Formal topology-based methods analyze shapes using structures such as Reeb graphs, persistent homology, and Morse--Smale complexes \cite{hilaga2001topology,edelsbrunner2002topological,gyulassy2007efficient}. Hi-TOPS is technically distinct from these methods: it uses ``topology-aware'' only to describe structure-level connectivity and boundary-relevant geometric evidence.

\section{Method}
\label{sec:method}

\subsection{Structural-Scale Mismatch}
\label{subsect:mismatch}
\mypara{Problem Statement}
Given an input triangle mesh $\mathcal{M}=(\mathcal{V},\mathcal{F})$ normalized into a canonical bounding volume, our goal is to produce a part decomposition
$\mathcal{S}=\{\mathcal{M}_k\}_{k=1}^{K}$, where $\mathcal{M}_k=(\mathcal{V}_k,\mathcal{F}_k)$ and
$\bigcup_{k=1}^{K}\mathcal{F}_k=\mathcal{F}$ with $\mathcal{F}_k\cap\mathcal{F}_{k'}=\emptyset$ for $k\neq k'$.
Instead of directly predicting per-face labels, we formulate decomposition as a \emph{structural partitioning} process driven by a volumetric representation.

\mypara{Structural-Scale Decomposition}
We first discretize the normalized shape volume at a set of resolutions $r\in R $ (e.g., $\{16,32,64\}$), yielding voxel grids
$\mathcal{V}^{(r)}=\{v_i^{(r)}\}_{i=1}^{r^3}$ with voxel size $d_{v}^{(r)}=1/r$.
We denote the set of surface-intersecting voxels as
\begin{equation}
\Omega^{(r)}=\{\, i \mid v_i^{(r)}\cap \mathcal{M}\neq \emptyset \,\}.
\end{equation}
The induced structure exhibits clear scale-dependent behavior (Fig.~\ref{fig:res_comp}).
At low resolution (e.g., $r{=}16$), $\Omega^{(r)}$ provides stable coarse evidence with body-level layout for smooth-region boundaries, though thin joints may be merged.
Across moderate resolutions ($r{=}32$ and $64$), the grids provide complementary structural evidence that finer meso-scales better expose articulations, narrow joints, and thin structures. Because these regions are observed across a limited range of meso scales, they remain structurally stable while avoiding the fragmentation often introduced by overly fine voxelization.

This motivates us to seek a \textit{hierarchical structural representation} that serves as this bridge to balance abstraction with fidelity, when their resolutions lie within an appropriate \textit{meso-scale range}, rather than a single fixed scale.
Naturally, the informative meso-scale range is also shape-dependent with greater geometric complexity, e.g., denser tessellation and richer thin structures.
Overall, a hierarchy of moderate resolutions, such as $r\in\{16,32,64\}$, can cover coarse semantic layout, meso-level articulation evidence, and boundary-sensitive details within a compact operating range.
Although very high-resolution representations, such as $128^3$ or $256^3$, may capture finer local details, the additional structural benefit is often marginal compared with the introduced computational overhead.

\mypara{Geometric Intrinsic Cues}
Given an operating resolution $r$ (Fig.~\ref{fig:res_comp}), the voxelized surface support $\Omega^{(r)}$ provides a stable structural scaffold for decomposition. 
To further probe \emph{where} boundaries should occur and \emph{how} local parts are related within this scaffold, we augment each active voxel with a set of geometric intrinsic cues that are commonly used to characterize local shape transitions.
For each active voxel $v_i^{(r)}$ with center $\mathbf{x}_i^{(r)}$, we collect the set of surface samples falling inside it, denoted as $\mathcal{P}_i^{(r)}\subset\mathcal{M}$, and define a cue vector
\begin{equation}
\boldsymbol{\phi}_i^{(r)}=\Big[\phi_{i,\mathrm{curv}}^{(r)},\ \phi_{i,\mathrm{norm}}^{(r)},\ \phi_{i,\mathrm{den}}^{(r)}\Big]\in\mathbb{R}^3,
\end{equation}
where $\phi_{i,\mathrm{curv}}^{(r)}$ measures local curvature magnitude, $\phi_{i,\mathrm{norm}}^{(r)}$ captures normal inconsistency, and $\phi_{i,\mathrm{den}}^{(r)}$ encodes local sampling density.

As shown in Fig.~\ref{fig:res_comp} (top panels), these intrinsic cues are strongly scale-dependent: at low resolution, they show high contrast and clearly separate boundary-adjacent regions from smooth interiors, while at higher resolution, they become increasingly uniform as local neighborhoods shrink, weakening curvature/normal contrast and reducing density discriminability.
Overall, this observation highlights two limitations: (\emph{i}) intrinsic cues at any \emph{single} scale cannot consistently capture both coarse structural contrast and fine articulation details, and (\emph{ii}) any \emph{single} cue in isolation remains insufficient to capture semantic part boundaries, which require integrating complementary evidence within a coherent structural context.

\begin{figure*}[t]
    \centering
    \includegraphics[width=0.95\textwidth]{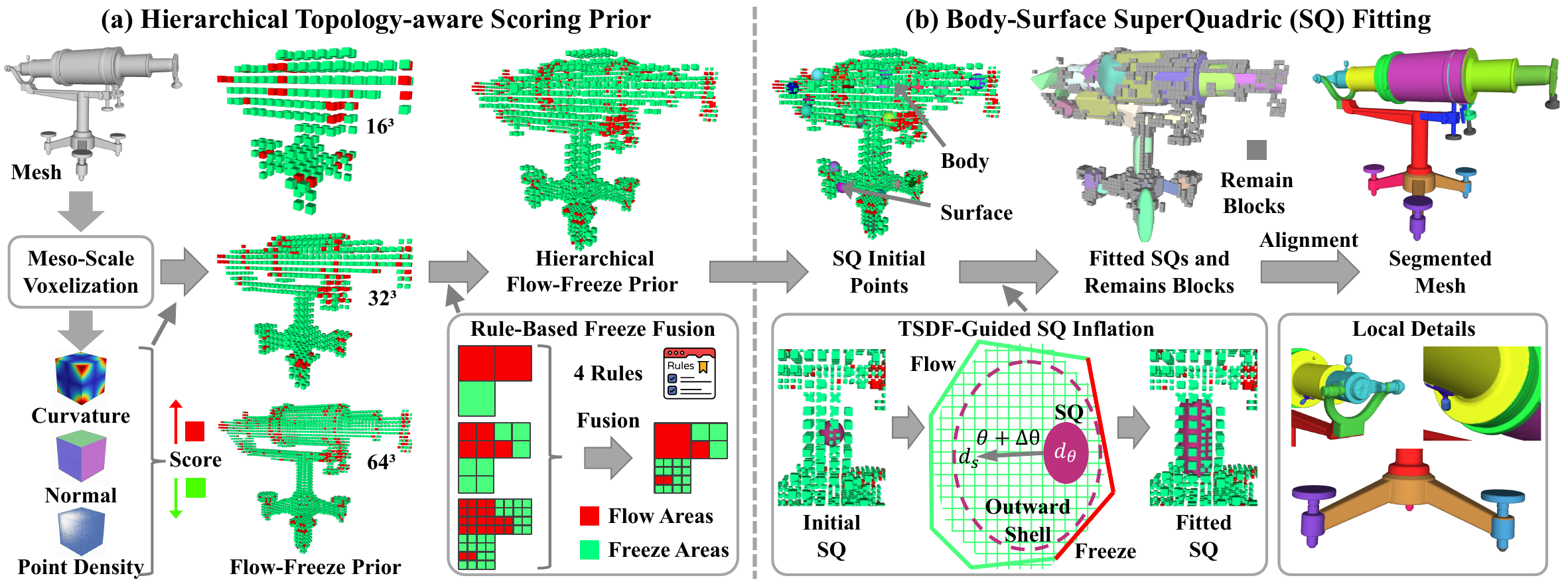}
    \caption{
    \textbf{Overview of the Hi-TOPS pipeline.}
    The input mesh is voxelized at multiple meso-scales to compute intrinsic cues and construct a hierarchical Flow-Freeze prior.
    This prior guides body-surface SQ initialization, TSDF-guided SQ inflation, and SQ-to-mesh assignment, producing part decomposition while preserving local details.
    }
    \label{fig:tops_score_ff}
\end{figure*}

\subsection{Hierarchical Topology-aware Scoring Prior}
\label{subsect:tops_prior}

Based on the resolution analysis in Sec.~\ref{subsect:mismatch}, we adopt a meso-scale operating range ($r{=}\{16, 32, 64\}$) to balance structural abstraction and boundary-sensitive fidelity.
In practice, we voxelize the normalized mesh $\mathcal{M}$ within a fixed bounding volume by rasterizing surface triangles into a uniform grid~\cite{zhou2018open3d}, producing a surface-aligned voxel set $\Omega$ for subsequent reasoning.
On this meso-scale scaffold, we construct a \textbf{hierarchical topology-aware scoring prior} with Flow-Freeze areas: low-score \emph{Flow} regions enable primitive expansion, while high-score \emph{Freeze} regions block cross-joint intrusion and preserve articulation boundaries.

\mypara{Topology-aware Score}
Building on the intrinsic cues $\phi_{i,\mathrm{curv}}^{(r)}$, $\phi_{i,\mathrm{norm}}^{(r)}$, and $\phi_{i,\mathrm{den}}^{(r)}$ in Sec.~\ref{subsect:mismatch}, we construct a \emph{topology-aware score} to consolidate boundary evidence into a single scalar field over the active voxel $v_i^{(r)}$ in set $\Omega^{(r)}$.
We further expand curvature into three statistics: curvature mean $\mu_i^{\kappa}$, curvature deviation $\sigma_i^{\kappa}$, and neighborhood curvature difference $\Delta_i^{\kappa}$.
Then, these terms are weighted by $\boldsymbol{\lambda}=\{\lambda_1,\lambda_2,\lambda_3,\lambda_4\}$ and modulated by a square-root density factor:
\begin{equation}
S_B(i)
\;=\;
\Big(\lambda_1\,\mu_i^{\kappa}
\;+\;
\lambda_2\,\sigma_i^{\kappa}
\;+\;
\lambda_3\,\Delta_i^{\kappa}
\;+\;
\lambda_4\,\phi_{i,\mathrm{norm}}^{(r)}\Big)
\cdot
\sqrt{\phi_{i,\mathrm{den}}^{(r)}}
\label{eq:SB}
\end{equation}
We refer to $S_B:\Omega^{(r)}\!\to\!\mathbb{R}_{\ge 0}$ as the \emph{structural field}: a scalar score defined on the active voxel set, whose top-$\alpha$ level set (Eq.~\ref{eq:freeze_prior}) acts as a geometry-driven proxy for part boundaries. Geometric Cue details are present in \textbf{Appendix.B}.

The boundary-sensitive score integrates curvature complexity, normal inconsistency, and density reliability into a \emph{topology-aware signal}.
Curvature and normal terms highlight bending transitions and articulation-like orientation changes, while density modulation suppresses unstable responses from sparsely supported regions.
Thus, $S_B$ provides a stable proxy for distinguishing boundary-critical zones, separating expandable support regions from boundaries.

\mypara{Flow-freeze Prior}
\label{sec:ffprior}
Given the topology-aware score $\{S_B(i)\}_{i\in\Omega^{(r)}}$, we directly induce a binary \textit{Flow-freeze} prior by selecting a fraction $\alpha\in(0,1)$ of the highest-scoring elements as \emph{Freeze}.
Let $\tau_\alpha$ denote the $(1-\alpha)$-quantile of $\{S_B(i)\}$, and define
\begin{align}
\mathcal{B}_{\mathrm{freeze}}
&=
\{\, i\in\Omega^{(r)} \mid S_B(i)\ge \tau_\alpha \,\}, \label{eq:freeze_prior}\\
\mathcal{B}_{\mathrm{flow}}
&=
\{\, i\in\Omega^{(r)} \mid S_B(i)< \tau_\alpha \,\}.
\label{eq:flow_freeze_prior}
\end{align}
This partition assigns explicit semantic roles to the voxelized structure.
Blocks in $\mathcal{B}_{\mathrm{flow}}$ correspond to low-complexity, volumetrically coherent regions that are suitable for forming stable part interiors; they serve as expandable support volumes where primitive growth and region merging are promoted.
In contrast, blocks in $\mathcal{B}_{\mathrm{freeze}}$ concentrate high-complexity evidence, typically arising near joints, thin attachments, and sharp structural transitions, and suppress cross-boundary propagation.
As a result, the Flow-Freeze prior enforces a decomposition dynamics that \emph{flows} through structurally consistent body regions while \emph{freezing} at articulation bottlenecks, preventing part leakage and producing geometric-consistent boundaries.

\mypara{Rule-based Freeze Fusion}
We fuse the priors coarse-to-fine: homogeneous Flow blocks remain coarse, finer Freeze evidence triggers refinement, and coarse Freeze labels are retained only when supported by most hildren; sparse and backward-consistency cases are detailed in Appendix~E.1.

Overall, this fusion keeps smooth body regions expandable at coarse scales while adding finer boundary constraints only with consistent geometric evidence. The hierarchical Flow-Freeze prior combines coarse stability with fine boundary sensitivity, providing compact guidance for primitive fitting and part decomposition.

\subsection{Body-Surface SuperQuadric Fitting}
\label{subsect:decomp}

With the hierarchical Flow-Freeze prior (Sec.~\ref{sec:ffprior}), we obtain volumetric guidance that separates expandable support regions ($\mathcal{B}_{\mathrm{flow}}$) from boundary-critical areas ($\mathcal{B}_{\mathrm{freeze}}$).
However, this voxel prior cannot be directly used as final mesh segmentations, since voxel partitions are coarse and may not align with the surface connectivity of $\mathcal{M}$.
To obtain accurate surface-level part boundaries, we introduce \textbf{Body-Surface SuperQuadric Fitting} with TSDF guidance as a voxel-to-mesh refinement stage.

\mypara{Body-Surface SQ Initialization}
We use SQ as volumetric primitives to explain expandable Flow regions $\mathcal{B}_{\mathrm{flow}}$ while respecting boundary-preserving Freeze regions $\mathcal{B}_{\mathrm{freeze}}$.
A general SQ is defined by an 11D parameter vector
$\bm{\theta}=(a_x,a_y,a_z,\epsilon_1,\epsilon_2,\bm{\rho},\bm{c})$,
where $a_x,a_y,a_z\in\mathbb{R}^{+}$ are scales, 
$\epsilon_1,\epsilon_2\in[0,2]$ are shape exponents, 
$\bm{\rho}\in\mathbb{R}^{3}$ parameterizes the rotation, and 
$\bm{c}\in\mathbb{R}^{3}$ denotes the SQ center.
Given the coordinate
$\bar{\bm{x}}=R(\bm{\rho})^{\top}(\bm{x}-\bm{c})=(\bar{x},\bar{y},\bar{z})^{\top}$,
the SQ surface is
\begin{equation}
f_{\bm{\theta}}(\bm{x})
=
\left(
\left( \frac{\bar{x}}{a_x} \right)^{\frac{2}{\epsilon_2}}
+
\left( \frac{\bar{y}}{a_y} \right)^{\frac{2}{\epsilon_2}}
\right)^{\frac{\epsilon_2}{\epsilon_1}}
+
\left( \frac{\bar{z}}{a_z} \right)^{\frac{2}{\epsilon_1}}
=1 .
\label{eq:superquadrics}
\end{equation}

Since SQ fitting is initialization-sensitive, we initialize $\bm{\theta}_0$ with a hierarchical coarse-to-fine Body-Surface seeding scheme.
At coarser scales, the \emph{Body} strategy places seeds at Euclidean distance transform (EDT) maxima inside $\mathcal{B}_{\mathrm{flow}}$, capturing dominant volumetric interiors.
At finer scales, the complementary \emph{Surface} strategy places seeds at centroids of remaining connected regions to recover thin attachments, shallow structures, and residual boundary-adjacent components.
Although the two strategies are not tied to a single resolution, this hierarchical order first establishes robust body cores and then supplements missing surface details.
The resulting candidate SQs grow inside Flow regions and are constrained near Freeze regions, providing stable boundary conditions for subsequent TSDF-guided inflation.
Details are present in \textbf{Appendix.E.2}.

\mypara{TSDF-Guided SQ Inflation}
Starting from the initialized SQ parameter $\bm{\theta}_0$, we iteratively inflate each primitive by expanding an outward shell around its current boundary with voxelization distance $d_{step}$.
At iteration $t$, the shell samples $\mathcal{P}_{\mathrm{shell}}^{t}$ are used to test whether the SQ can safely grow nearby.
We refine the SQ by minimizing the TSDF discrepancy between the SQ and outward shell:
\begin{equation}
\min_{\bm{\theta}\in[\bm{\theta}_{\min},\bm{\theta}_{\max}]}
\sum_{\bm{p}\in\mathcal{P}_{\mathrm{shell}}^{t}}
w(\bm{p})
\left(
\tilde{d}_{\bm{\theta}}(\bm{p})
-
d_{\mathcal{M}}(\bm{p})
\right)^2,
\label{eq:sdf_sq_fit}
\end{equation}
where $d_{\mathcal{M}}$ is the clipped TSDF of outward shell, $\tilde{d}_{\bm{\theta}}$ is the SQ TSDF, and $w(\bm{p})$ balances surface alignment and exterior intrusion penalty.
We apply the optimized increment with a backtracked step size and accept it only when the updated shell remains in Flow without intruding into Freeze; implementation details are in \textbf{Appendix E.2}.

Overall, TSDF-guided SQ inflation grows each initialized seed into a geometric-consistent part by aligning the SQ with the mesh TSDF while restricting its expansion within Flow regions and stopping near Freeze boundaries.

\mypara{SQ-to-Mesh Assignment}
After SQ fitting, we obtain fitted SQs $\{Q_k\}_{k=1}^{K}$, each inducing a semantic voxel support $\mathcal{B}_k\subseteq\mathcal{B}_{\mathrm{flow}}$.
However, these voxel supports cannot directly define the final mesh segmentation, since they may leave remaining blocks and may not align with the surface connectivity of $\mathcal{M}$.
We first decompose $\mathcal{M}$ into curvature-aware edge primitives $\{\mathcal{E}_j\}_{j=1}^{A}$.
This is done by scoring shared edges using dihedral angles, cutting high-curvature edges above the $q{=}90$ quantile of the per-mesh distribution, and applying edge-aware flood-fill on the resulting face graph~\cite{chen2025dora}.
These edge primitives provide geometrically coherent surface units, but do not carry semantic part labels by themselves.

For each edge primitive $\mathcal{E}_j$, we transfer SQ labels from voxel space to mesh space in three steps.
First, for every face $f\in\mathcal{E}_j$, we query the nearest SQ support in $\bigcup_k\mathcal{B}_k$ using the face centroid and record its closest SQ label.
Second, we assign $\mathcal{E}_j$ the distance-weighted majority SQ label among its faces, denoted as $k_j^{*}$.
This converts noisy face-level nearest-neighbour labels into a stable edge-primitive-level label.
Third, edge primitives sharing the same SQ label are merged to form the final mesh parts:
\begin{equation}
\mathcal{S}=\{\mathcal{M}_k\}_{k=1}^{K},\qquad
\mathcal{M}_k=\bigcup_{j:\,k^{*}_j=k}\mathcal{E}_j .
\label{eq:final_segmentation}
\end{equation}
Thus, SQ supports provide structural part semantics, while curvature-aware edge primitives keep the final seams aligned with mesh geometry.
Details are present in \textbf{Appendix.E.4}.

\begin{table}[t]
    \centering

    \caption{Quantitative comparison of segmentation results on PartObjaverse-Tiny, PartNet and HY3D-Bench datasets. The evaluation metric is \textbf{mIoU} ($\uparrow$). Learning-based methods run on an \textbf{RTX 5090 GPU}, while training-free methods run \textbf{single-threaded on an AMD Ryzen 9 5900X CPU}.} 
    \label{tab:segmentation_results}
    
    \Description{A table comparing the mIoU performance of EMS, MPS, PrimitiveAnything, $S^2$AM3D, SAMPart3D, PartSAM, Partfield, PointSAM and our method. Our method achieves the highest scores.}
    
    \scalebox{0.9}{
    \begin{tabular}{lcccccc}
        \toprule
        Method  & Train & PartObj & PartNet & HY3D-B & Time\\
        \midrule
        
        EMS         & \ding{56} & 14.24            & 11.21            &  16.96   & 1.304s \\
        MPS         & \ding{56} & 23.72           & 23.89            & 25.25    & 	9.207s\\
        \midrule
        PrimAny     &\ding{52}  & 26.54            & 22.81            & 26.40    & 36.92s\\
        $S^2$AM3D   &\ding{52}  & 30.64            & 29.77            & 22.2     & 5.235s \\
        SAMPart3D   & \ding{52} & 46.74            & 29.48            & 43.9     & 29.33s \\
        PartSAM     & \ding{52} & 49.15            & 30.16            & \textbf{55.13} & 20.05s  \\
        Partfield   & \ding{52} & \textbf{66.77}    & \underline{45.57} & 44.56     & 65.41s   \\
        PointSAM    & \ding{52} & 43.27              & 27.66            & 32.94    & 42.20s \\
        \midrule
        \textbf{Hi-TOPS} & \ding{56} & \underline{51.63} & \textbf{55.87} & \underline{47.22} & 129.5s    \\ 
        
        \bottomrule
    \end{tabular}}
\end{table}

\section{Experiments}
\label{sec:exp}

\subsection{Implementation  Details}
\mypara{Datasets}
We evaluate Hi-TOPS on three public part-annotated datasets: PartObjaverse-Tiny~\cite{sampart3d_yang2024}, PartNet~\cite{partnet_2019} and HY3D-Bench ~\cite{hy3d_2026}. PartObjaverse-Tiny provides 200 diverse instances with part-level labels, enabling evaluation under significant shape variation. PartNet contains 100 instances with frequent articulations and thin attachments, including \textsc{Chair}, \textsc{Table}, \textsc{Lamp}, \textsc{Bed} and \textsc{StorageFurniture}. We randomly sample 200 meshes from HY3D-Bench as a generalization set.

\mypara{Evaluation Metrics}
We evaluate only the mesh-level segmentation rather than surface-faithful primitive fitting, reporting mean IoU (mIoU) under the standard protocol with greedy label matching.
To assess structural correctness around articulations, we additionally compute Rand Index (RI), Variation of Information (VoI), and Segmentation Covering (SC). 
We compare Hi-TOPS against primitive-based abstraction methods~\cite{primAny, mps_2023, ems_2022} (converted to surface partitions) and recent 3D part segmentation models~\cite{sampart3d_yang2024, s2am3d_su2025, parfield_2025, partsam_zhu2025, pointsam_zhou2025}.
Details are present in \textbf{Appendix.A}.

\begin{figure*}[ht]
    \centering
    \includegraphics[width=0.96\textwidth]{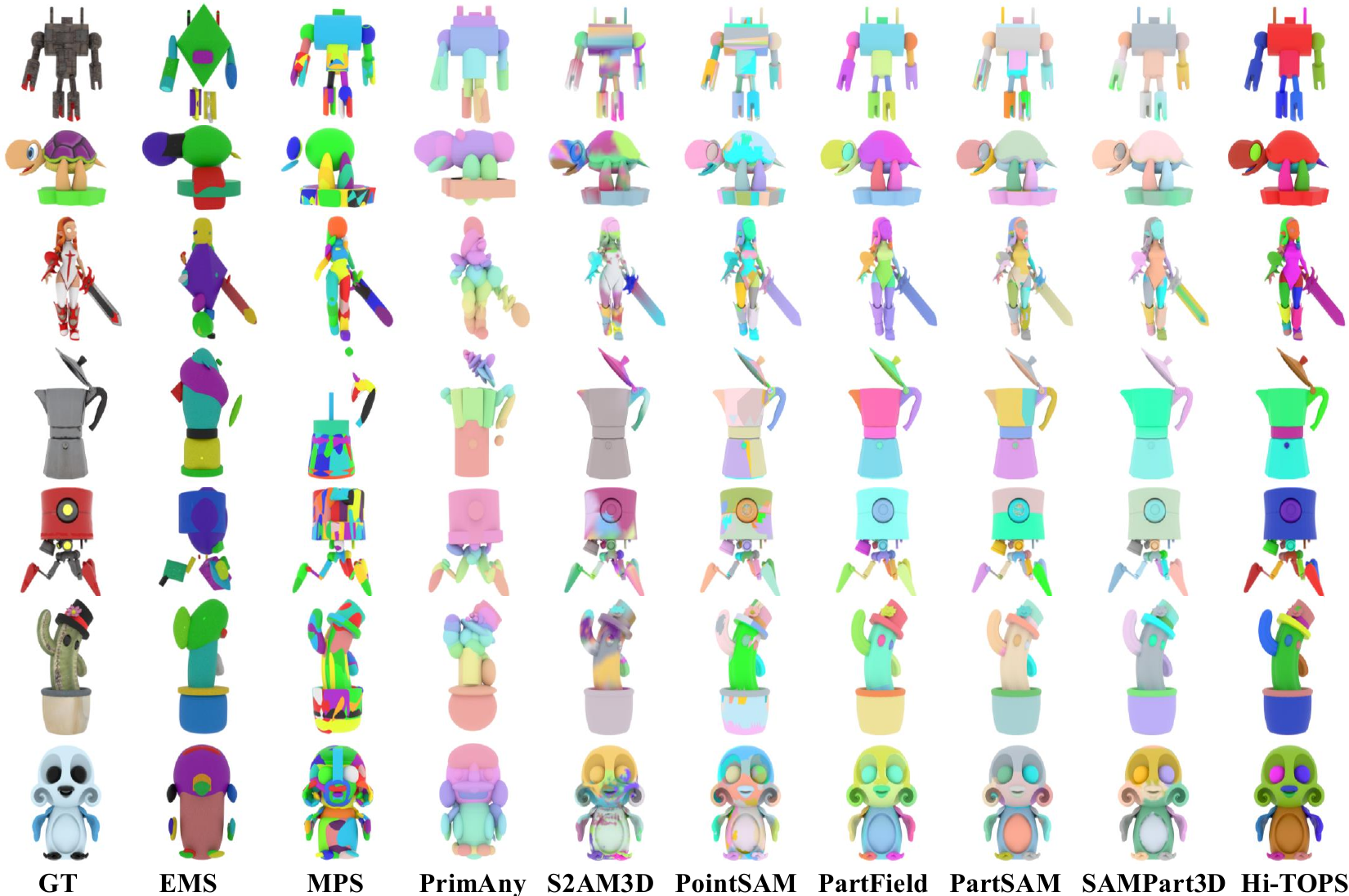}
    \caption{Qualitative comparison of 3D part segmentation on diverse meshes. We visualize segmentation results of primitive-based baselines, learning-based methods, and our Hi-TOPS on shapes with varying articulation complexity and component scale. }
    \label{fig:qualitive}
\end{figure*}

\mypara{Setup}
We use a single parameter setting for Hi-TOPS across every dataset and every shape: the resolution set is fixed to $\mathcal{R}{=}\{16,32,64\}$, the per-face barycentric sample count to $N_\mathrm{face}{=}36$, the four cue weights to $\boldsymbol{\lambda}{=}(1,1,1,1)$, the Freeze ratio to $\alpha{=}0.15$, and the dihedral edge-cut threshold to the $90$-th quantile of the per-mesh dihedral distribution ($q{=}90$). These five settings dominate the structural behaviour of the pipeline; all other constants are fixed defaults listed in \textbf{Appendix.A}, and no per-shape or per-dataset tuning is applied.

\subsection{Main Results}
\mypara{Quantitative Evaluation}
We evaluate whether geometry-only meso-scale topology can support competitive part segmentation without semantic training or 2D priors.
Table~\ref{tab:segmentation_results} reports mIoU on three datasets.
Hi-TOPS clearly outperforms primitive-based methods, showing that primitive expressiveness alone is insufficient without topology-aware boundary control.
Against learning-based models, Hi-TOPS is most consistent across datasets (cross-dataset std 3.5 vs ~10.5), and strongest on PartNet.
Although it is slightly behind PartField on PartObjaverse-Tiny and PartSAM on HY3D-Bench, these methods do not consistently dominate across all datasets, likely due to their dependence on training coverage, learned semantics, and category priors.
By relying only on intrinsic geometry and meso-scale topology, Hi-TOPS achieves more consistent cross-dataset performance. Detailed analysis is provided in \textbf{Fig.~\ref{fig:deep_analysis} and Appendix.C}.
Since Hi-TOPS currently runs on a single-thread CPU, its runtime still has substantial optimization potential.
Table~\ref{tab:prim_seg_comparison_merged} extends the comparison to structural metrics. Hi-TOPS clearly outperforms all primitive baselines and is best on PartNet in VoI and SC, while PartField leads on PartObjaverse-Tiny. Its cross-dataset variation is the smallest (SC 0.544$\to$0.532 vs.\ 0.541$\to$0.333 for SAMPart3D), indicating that structural fidelity is preserved without semantic supervision.

\begin{table}[t]
    \centering
    \caption{Structural metrics comparison on \textbf{PartObjaverse-Tiny} and \textbf{PartNet} against both primitive-based and learning-based baselines. Metrics include Rand Index (RI), Variation of Information (VoI), and Segmentation Covering (SC).}
    \label{tab:prim_seg_comparison_merged}
    \setlength{\tabcolsep}{4.5pt} 
     \scalebox{0.9}{\begin{tabular}{l c ccc c ccc}
        \toprule
        & & \multicolumn{3}{c}{\textbf{PartObjaverse-Tiny}} & \phantom{a} & \multicolumn{3}{c}{\textbf{PartNet}} \\
        \cmidrule{3-5} \cmidrule{7-9}
        Method & Train & RI $\uparrow$ & VoI $\downarrow$ & SC $\uparrow$ && RI $\uparrow$ & VoI $\downarrow$ & SC $\uparrow$ \\
        \midrule
        MPS     & \ding{56} & 0.719 & 2.37 & 0.295 && 0.734 & 2.216 & 0.338 \\
        EMS     & \ding{56} & 0.373 & 1.770 & 0.283 && 0.372 & 1.897 & 0.277 \\
        PrimAny & \ding{52} & 0.744 & 2.453 & 0.292 && 0.746 & 2.561 & 0.273 \\
        \midrule
        SAMPart3D   & \ding{52} & \textbf{0.818} & 1.300 & 0.541  && 0.574 & 1.949 & 0.333  \\
        PartSAM     & \ding{52} & 0.772 & 1.563 & 0.481  &&  0.791 & 1.814 & 0.427 \\
        Partfield   & \ding{52} & \textbf{0.815} & \textbf{1.088} & \textbf{0.548}  && 0.837 & \underline{1.625} & \underline{0.493}   \\
        \midrule
        Ours    & \ding{56} & 0.778 & \underline{1.237} & \underline{0.544} && \underline{0.796} & \textbf{1.444} & \textbf{0.532} \\
        \bottomrule
    \end{tabular}}
\end{table}

\begin{figure*}[ht]
    \centering
    \includegraphics[width=0.95\textwidth]{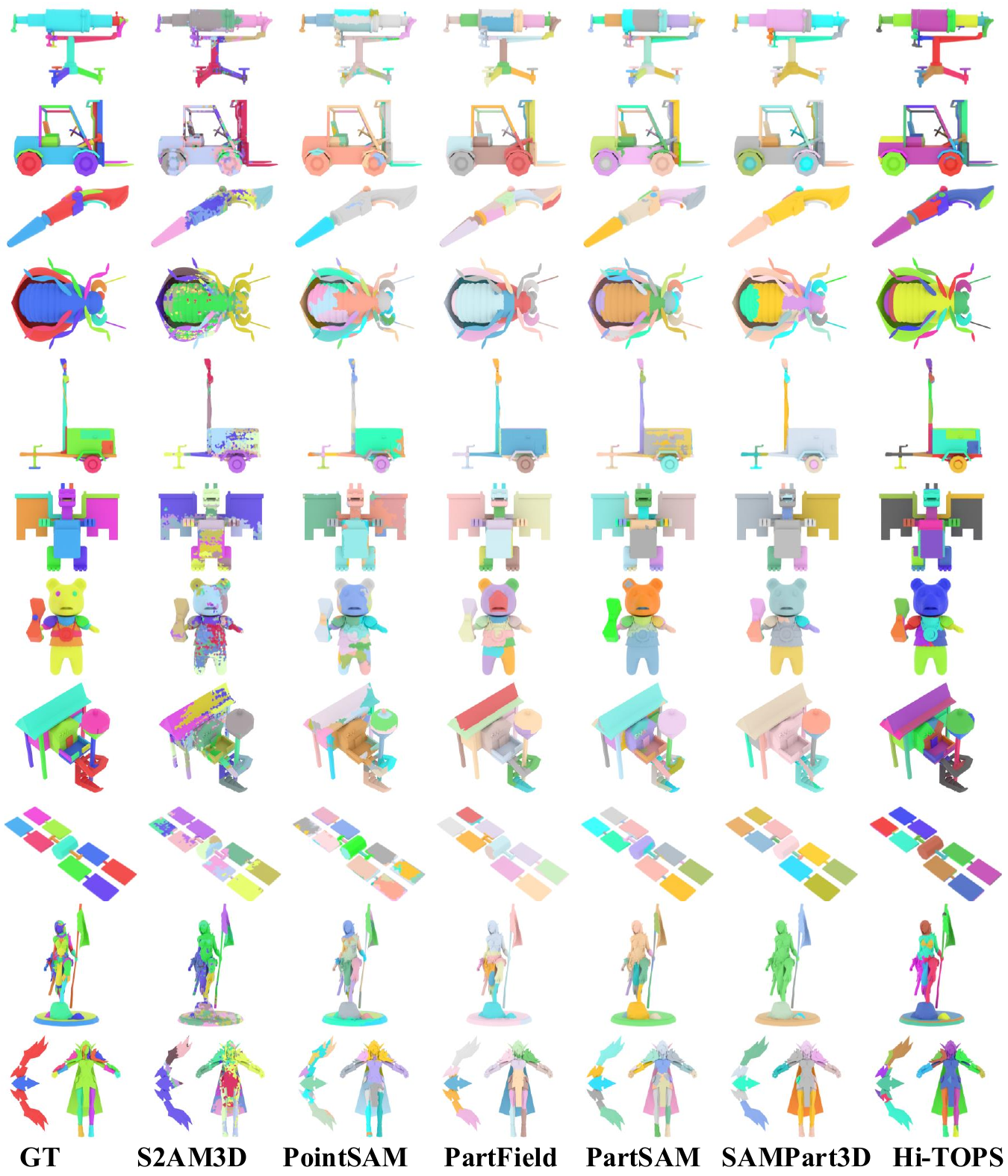}
    \caption{More qualitative comparison case results. Our method demonstrates competitive segmentation performance with topologically faithful structure.}
    \label{fig:qualitive2}
\end{figure*}

\mypara{Qualitative Evaluation}
Fig.~\ref{fig:qualitive} and~\ref{fig:qualitive2} compare Hi-TOPS with primitive-based and learning-based baselines.
Primitive methods often under- or overfit, missing thin structures, crossing articulations, or producing distorted primitives on complex shapes. Hi-TOPS instead uses Flow--Freeze prior to guide SQ growth and obtain stable, boundary-aligned supports. Learning-based methods may merge unseen parts or produce fragmented, cluster-dependent results. Overall, Hi-TOPS provides robust decomposition without semantic supervision; additional cases are shown in \textbf{Appendix~F}.

\subsection{Ablation Study}
\label{subsec:ab_study}
We conduct ablations at two levels.
Table~\ref{tab:ablation_tops} isolates component-level design choices under a single-scale setting.
Removing Body initialization, Freeze constraints, or any geometric cue degrades the results, confirming the importance of volumetric core discovery, boundary restriction, and complementary curvature/normal/density.
The Flow-Freeze threshold study further shows that an overly small or large Freeze ratio weakens the balance between expansion and boundary preservation.

Table~\ref{tab:prior_adap_sdf_ablation} studies the system-level effects of TSDF guidance, Flow-Freeze constraints, and hierarchical resolutions.
TSDF alone performs weakly without Freeze constraints, while Freeze without TSDF improves structure but loses surface consistency. Removing TSDF also increases runtime at matched resolution. (\textbf{Appendix D.3})
Combining both gives the strongest single-scale results, with $r{=}64$ performing best among single resolutions.
The hierarchical setting further improves performance: $\{32,64,128\}$ gives the highest accuracy, 
whereas $\{16,32,64\}$ achieves nearly comparable results with much lower runtime.
This confirms that a compact meso-scale hierarchy provides the best accuracy-efficiency trade-off.

\begin{table}[t]
    \centering
    \caption{Ablation study with single-scale on a subset of PartObjaverse-Tiny.
    We evaluate the impact of (i) superquadric composition, (ii) Flow--Freeze threshold selection, and (iii) geometric cues on RI, VoI, SC, and mIoU.}
    \label{tab:ablation_tops}
    
    \scalebox{0.93}{
    \begin{tabular}{llcccc}
        \toprule
        Category & Variant & RI $\uparrow$ & VoI $\downarrow$ & SC $\uparrow$ & mIoU $\uparrow$ \\
        \midrule

        \multirow{2}{*}{\textbf{SQ Comp.}} 
          & w/o Body       & 0.731 & 2.360 & 0.307 & 31.94 \\
          & w/o Freeze     & 0.731 & 2.127 & 0.358 & 33.03 \\ 
        \midrule
        
        \multirow{2}{*}{\shortstack{\textbf{Flow-Freeze}\\\textbf{Threshold}}} 
          & Top 5\%        & 0.724 & 2.352 & 0.316 & 31.07 \\
          & Top 25\%       & 0.714 & 2.296 & 0.316 & 30.50 \\
        \midrule

        \multirow{3}{*}{\shortstack{\textbf{Geometric}\\\textbf{Cues}}} 
          & w/o Curvature  & \textbf{0.735} & 2.217 & 0.332 & 31.48 \\ 
          & w/o Normal     & 0.719 & 2.317 & 0.326 & 30.89 \\ 
          & w/o Density    & 0.734 & 2.192 & 0.332 & 31.93 \\ 
        \midrule

        \multicolumn{1}{c}{\textbf{w/o TSDF}} & Res=32 
          & 0.715 & \textbf{1.820} & \textbf{0.394} & \textbf{34.67} \\

        \bottomrule
    \end{tabular}}
\end{table}

\begin{table}[t]
    \centering
    \caption{Ablation study on a subset of PartObjaverse-Tiny.
    We compare single-scale and hierarchical settings with different combinations of TSDF and Flow-Freeze constraints. Metrics include RI, VoI, SC, mIoU, and runtime.}
    \label{tab:prior_adap_sdf_ablation}
    \scalebox{0.88}{
    \begin{tabular}{ccccccc}
        \toprule
        Setting & Resolution & RI $\uparrow$ & VoI $\downarrow$ & SC $\uparrow$ & mIoU $\uparrow$ & Time $\downarrow$ \\
        \midrule

        \multirow{4}{*}{\shortstack{Single Scale\\ w TSDF \\ w/o Freeze}}
            & 16  & 0.623 & 1.507 & 0.424 & 27.91 & 18.8s \\
            & 32  & 0.547 & 1.567 & 0.380 & 25.52 & 19.9s \\
            & 64  & 0.568 & 1.555 & 0.392 & 27.77 & 41.7s \\
            & 128 & 0.536 & 1.595 & 0.387 & 26.62 & 162.2s \\
        \midrule

        \multirow{4}{*}{\shortstack{Single Scale\\ w/o TSDF \\ w Freeze}}
            & 16  & 0.716 & 2.132 & 0.348 & 30.68 & 38.1s \\
            & 32  & 0.715 & 1.820 & 0.394 & 34.67 & 44.5s \\
            & 64  & 0.672 & 1.774 & 0.386 & 32.52 & 101.6s \\
            & 128 & 0.675 & 1.618 & 0.412 & 31.06 & 234.1s \\
        \midrule

        \multirow{4}{*}{\shortstack{Single Scale\\ w TSDF \\ w Freeze}}
            & 16  & 0.743 & 1.164 & 0.563 & 43.95 & 19.6s \\
            & 32  & 0.754 & 1.231 & 0.537 & 47.29 & 31.5s \\
            & 64  & 0.757 & 1.225 & 0.551 & 49.95 & 60.4s \\
            & 128 & 0.709 & 1.279 & 0.519 & 46.80 & 229.1s \\
        \midrule

        \multirow{1}{*}{\shortstack{w/o TSDF}}
            & $\{16,32,64\}$ & 0.749 & 1.356 & 0.506 & 47.17 & 238.4s \\
        \midrule

        \multirow{3}{*}{\shortstack{Ours\\Hierarchical\\w TSDF}}
            & $\{8,16,32\}$    & 0.748 & 1.126 & 0.559 & 45.98 & 95.4s \\
            & $\{32,64,128\}$  & \textbf{0.830} & \textbf{1.101} & \textbf{0.605} & \textbf{57.66} & 471.3s \\
            & $\{16,32,64\}$   & \underline{0.826} & \underline{1.144} & \underline{0.589} & \underline{56.56} & 132.1s \\

        \bottomrule
    \end{tabular}}
\end{table}

\subsection{Additional Experiments}
\begin{figure*}[ht]
    \centering
    \includegraphics[width=0.7\textwidth]{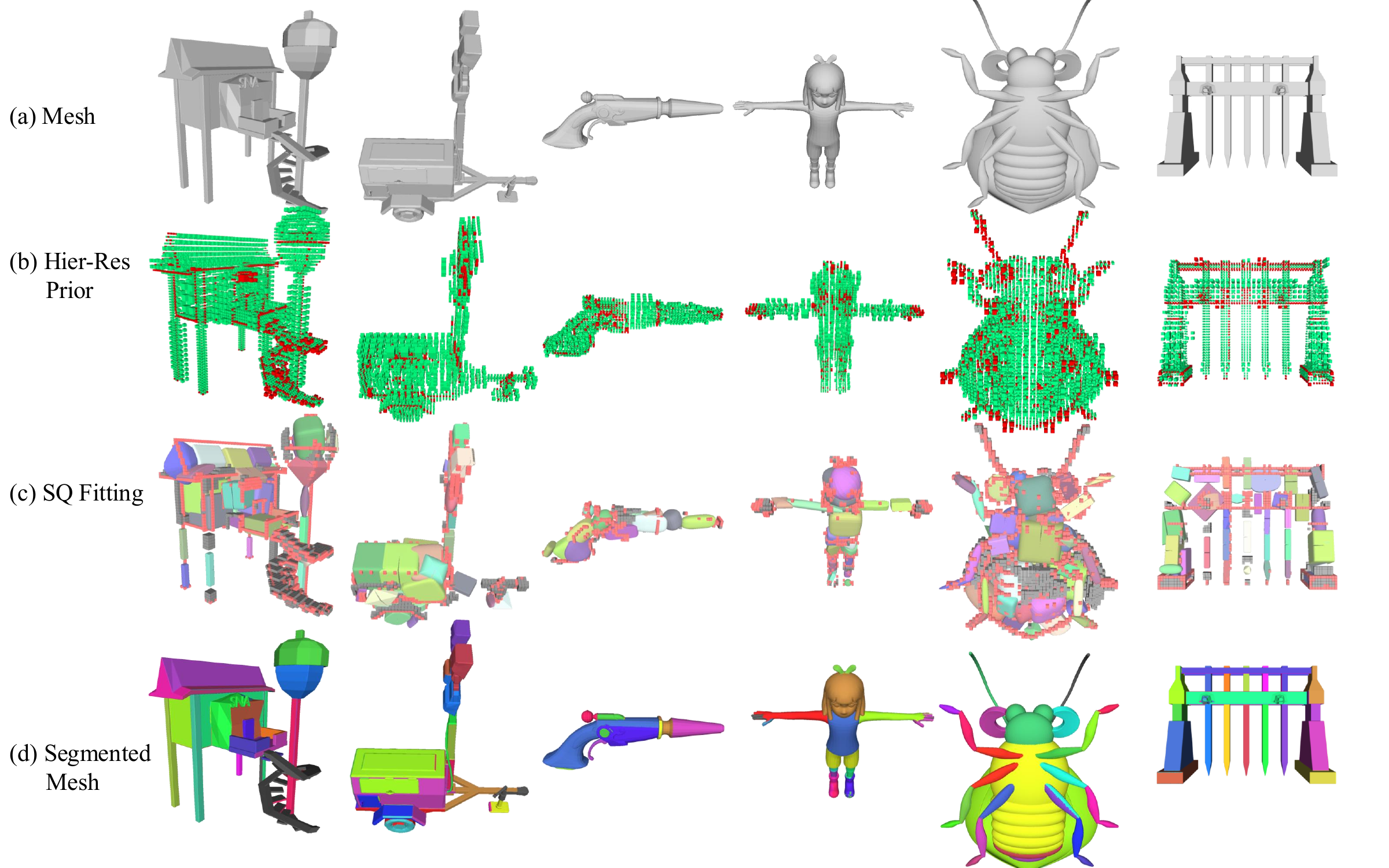}
    \caption{From top to bottom, we show (a) the input mesh, (b) the hierarchical-resolution prior, (c) the superquadric (SQ) fitting results guided by the prior, and (d) the final segmented mesh. The hierarchical prior provides meso-scale structure for primitive fitting and part decompositions across diverse geometries.}
    \label{fig:prior_vis}
    \vspace{10pt}
\end{figure*}

\begin{figure*}[h]
    \centering

    \begin{minipage}{0.35\textwidth}
        \centering
        \includegraphics[width=\linewidth]{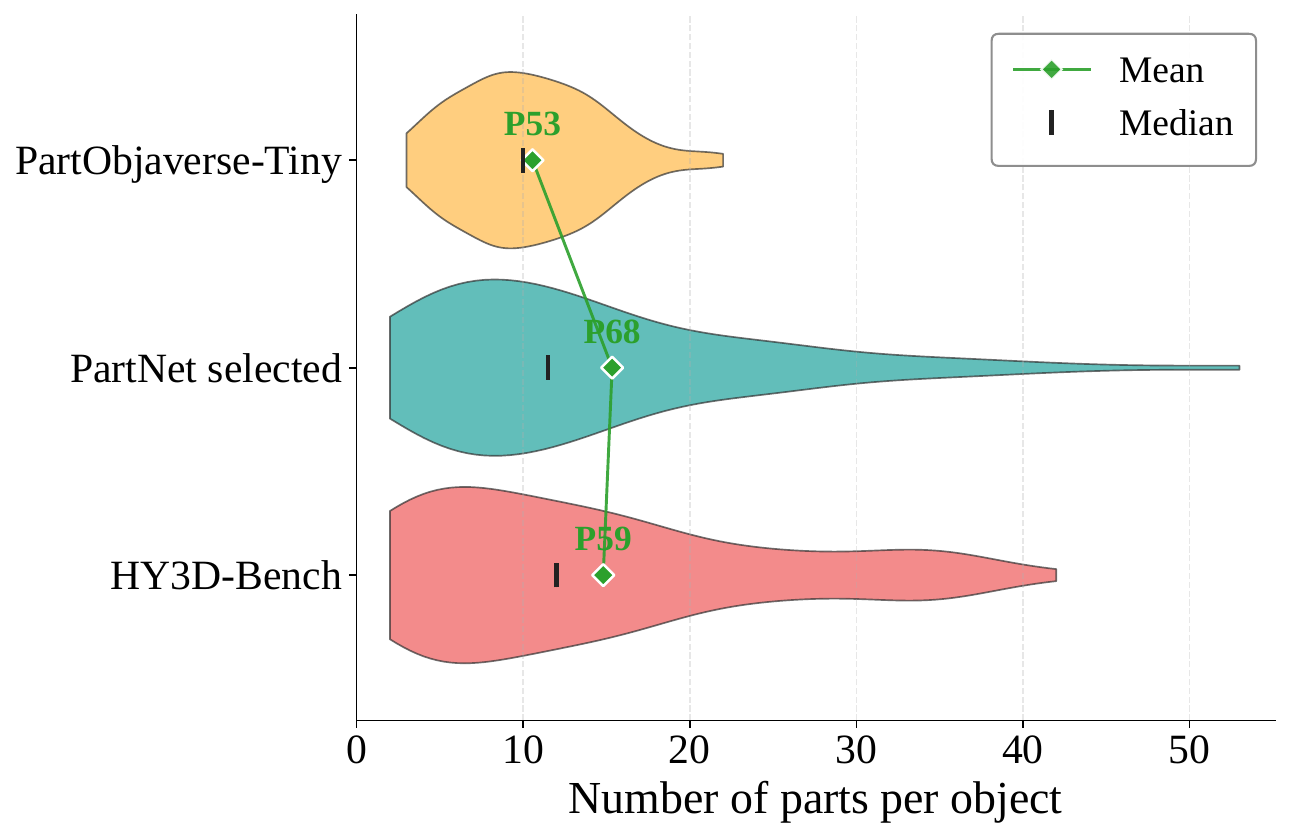}
        \vspace{-4pt}
        {\small (a) Part-count distribution}
    \end{minipage}
    \hfill
    \begin{minipage}{0.35\textwidth}
        \centering
        \includegraphics[width=\linewidth]{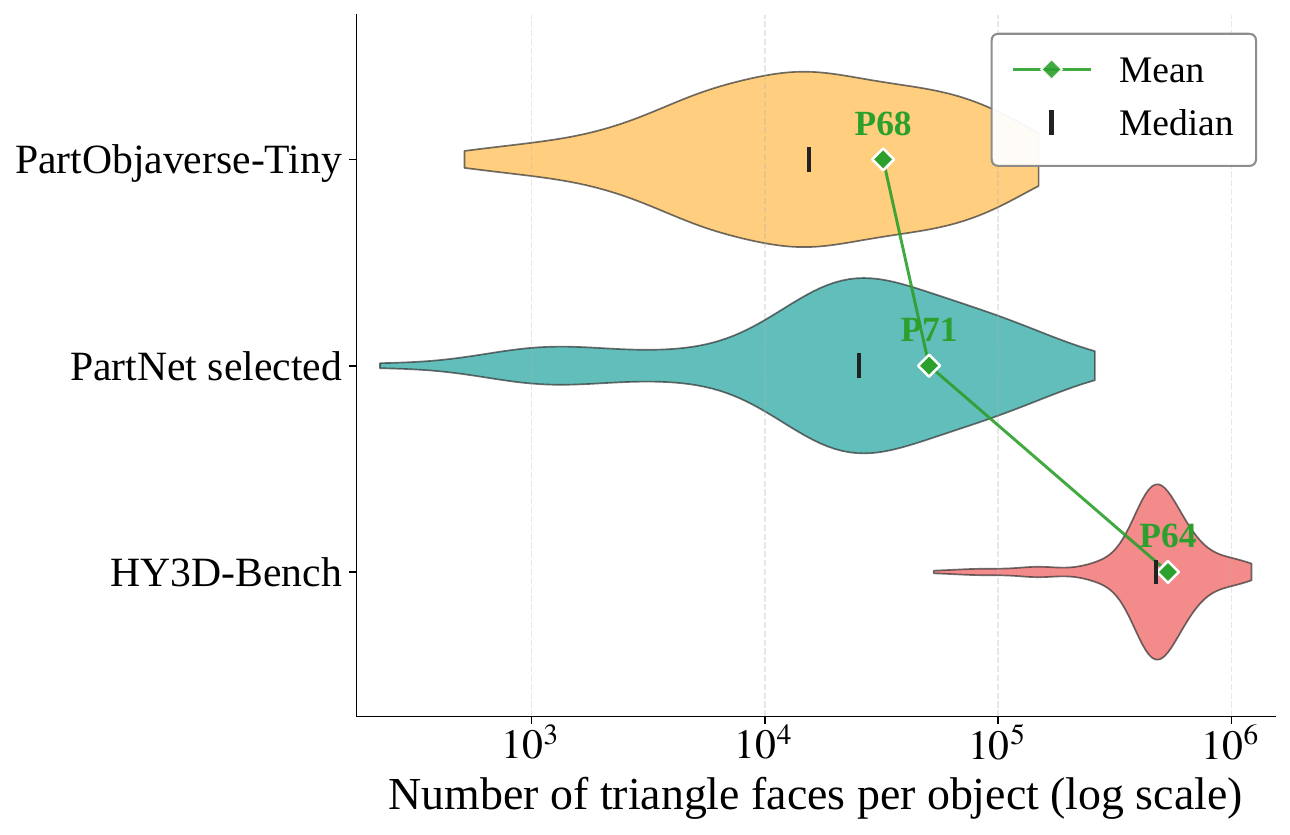}
        \vspace{-4pt}
        {\small (b) Face-count distribution}
    \end{minipage}
    \hfill
    \begin{minipage}{0.25\textwidth}
        \centering
        \begin{minipage}[t]{0.45\linewidth}
            \centering
            \includegraphics[width=\linewidth]{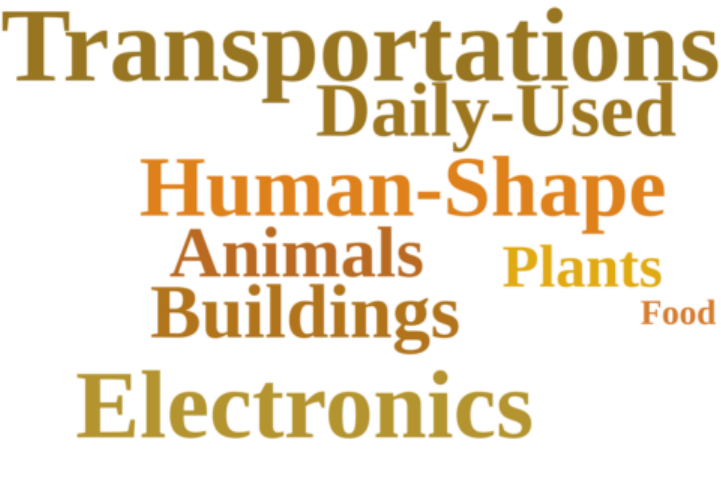}
            \vspace{-4pt}
        \end{minipage}
        \vfill
        \begin{minipage}{0.45\linewidth}
            \centering
            \includegraphics[width=\linewidth]{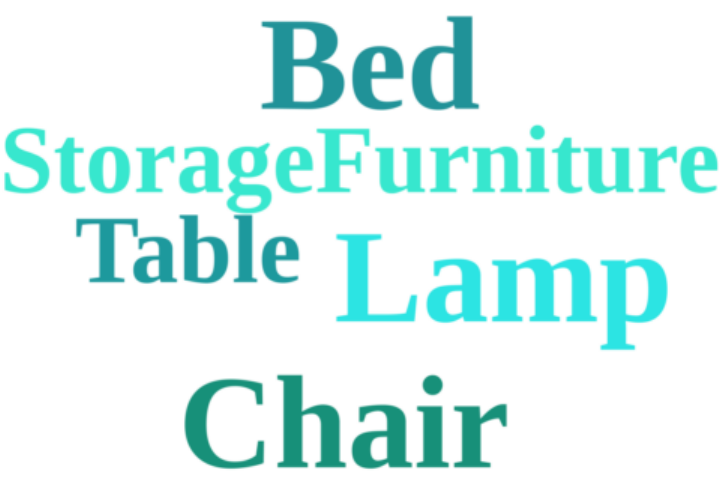}
            \vspace{-4pt}
        \end{minipage}
        \vfill
        \begin{minipage}{0.45\linewidth}
            \centering
            \includegraphics[width=\linewidth]{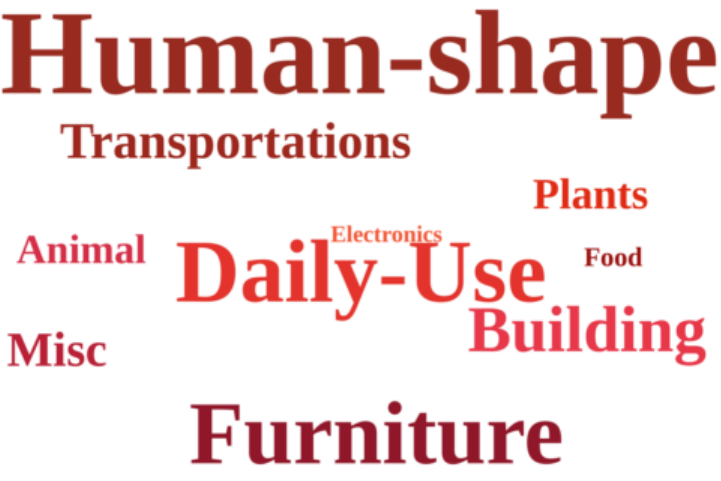}
            \vspace{-4pt}
        \end{minipage}

        \vspace{2pt}
        {\small (c) Category word clouds}
    \end{minipage}

    \caption{
    {Dataset-distribution analysis for PartSAM and PartField.}
    We compare part-count, face-count, and category distributions across PartNet selected, PartObjaverse-Tiny, and HY3D-Bench.
    PartObjaverse-Tiny is closer to PartField's training distribution, while HY3D-Bench is closer to PartSAM's training distribution, explaining their dataset-specific advantages in Table~\ref{tab:segmentation_results}.
    Differences in part numbers, mesh complexity, and category coverage further affect learning-based methods under distribution shift.
    By relying on geometry-driven structural cues instead of dataset-specific priors, Hi-TOPS adapts more consistently across datasets.
    }
    \label{fig:deep_analysis}
\end{figure*}

We provide additional visual results to complement the main quantitative and qualitative evaluations.
Figs.~\ref{fig:prior_vis} and \ref{fig:deep_analysis} provide additional pipeline, ablation, and dataset analyses, with detailed interpretations in their captions. 
Furthermore, \textbf{Appendix C} analyzes datasets in detail, \textbf{Appendix D} provides additional ablations and runtime analysis, and \textbf{Appendix F} shows more visualization cases.

\section{Conclusion and Limitation}
\label{sec:conclusion}

In this work, we introduce Hi-TOPS, a hierarchical topology-aware scoring prior for geometry-driven 3D part decomposition.
Hi-TOPS mitigates the structural-scale mismatch by aggregating intrinsic geometric cues within a compact meso-scale voxel hierarchy, where coarse scales preserve stable body-level support and finer scales recover articulation-sensitive boundaries.
The resulting Flow--Freeze prior guides Body-Surface SQ initialization, TSDF-guided SQ inflation, and SQ-to-mesh assignment, enabling volumetric primitives to be converted into clean mesh-level parts.
Extensive experiments show that Hi-TOPS outperforms primitive-based baselines and achieves competitive performance against recent learning-based segmentation models, while requiring no semantic supervision or 2D foundation priors.
These results demonstrate that meso-scale structural cues provide an effective basis for geometric-consistent, articulation-preserving 3D part decomposition.

Hi-TOPS works best when object parts can be represented by coherent volumetric cores separated by localized structural transitions.
Its local geometric cues may remain sensitive to tessellation and dense surface details, while  extremely concave geometry, highly non-convex parts, or very fine thin structures, the superquadric representation and finite voxel carrier may still underfit local details.
Future work may explore more flexible primitive families.
Finally, SQ fitting dominates the runtime of our current single-threaded CPU implementation; parallel voxelization, batched TSDF evaluation, and accelerated SQ fitting could substantially improve efficiency.

\clearpage
\bibliographystyle{ACM-Reference-Format}
\bibliography{tops_ref}

\clearpage

\begin{figure*}[ht]
    \centering
    \includegraphics[width=0.7\textwidth]{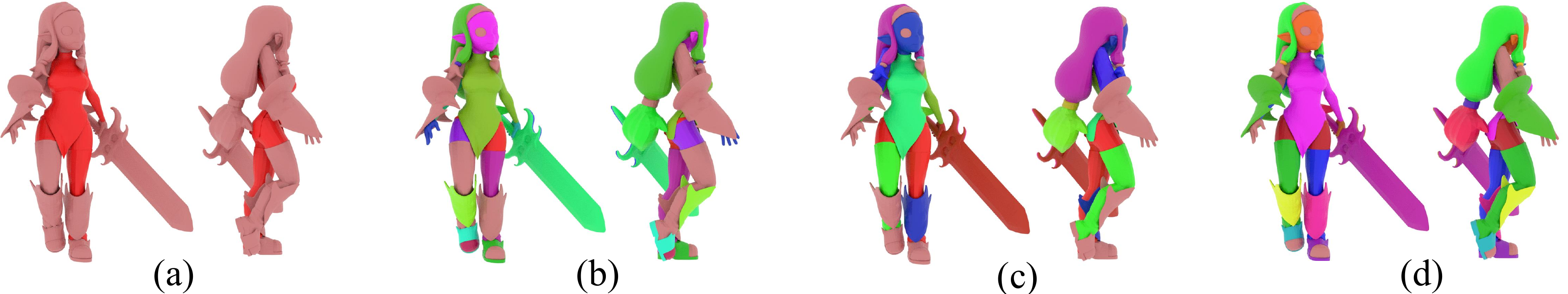}
    \caption{From left to right, we compare four settings from Table~\ref{tab:prior_adap_sdf_ablation}: 
    (a) single-scale w TSDF / w/o Freeze, 
    (b) single-scale w Freeze / w/o TSDF, 
    (c) hierarchical w Freeze / w/o TSDF, and 
    (d) full Hi-TOPS with hierarchical Flow--Freeze and TSDF-guided SQ fitting. 
    Flow-Freeze improves articulation separation, hierarchical fusion improves multi-scale structural coverage, and TSDF-guided fitting improves surface alignment, yielding the most coherent decomposition.}
    \label{fig:ablation_vis}
   
\end{figure*}

\clearpage

\clearpage


\appendix
\tableofcontents

\section*{\centering APPENDIX}
\addcontentsline{toc}{section}{Appendix}

\renewcommand{\thefigure}{A.\arabic{figure}}
\renewcommand{\thetable}{A.\arabic{table}}
\renewcommand{\thealgorithm}{A.\arabic{algorithm}}
\setcounter{figure}{0}
\setcounter{table}{0}
\setcounter{algorithm}{0}
\section{Implementation  Details}
\mypara{Datasets}
We evaluate Hi-TOPS on three public part-annotated datasets: PartObjaverse-Tiny~\cite{sampart3d_yang2024}, PartNet~\cite{partnet_2019} and HY3D-Bench ~\cite{hy3d_2026}. PartObjaverse-Tiny has 200 shapes and 8 categories, provides diverse object instances with part-level labels, enabling evaluation under significant shape variation. From PartNet, we focus on categories with frequent articulations and thin attachments, sampling 20 shapes from each of \textsc{Chair}, \textsc{Table}, \textsc{Lamp}, \textsc{Bed}, and \textsc{StorageFurniture}, resulting in 100 shapes in total. We randomly sample 200 meshes from HY3D-Bench as a generalization set. Following the category definitions of GObjaverse~\cite{richdreamer_2024_qiu} and PartObjaverse-Tiny~\cite{sampart3d_yang2024}, we categorize the 200 HY3D-Bench objects into 10 groups: Animal (11), Building (21), Furniture (33), Daily-Use (33), Electronics (5), Food (6), Human-Shape (50), Plants (12), Transportations (15), and Miscellaneous (14).

\mypara{Evaluation Metrics}
We evaluate the final mesh-level segmentation of Hi-Tops. 
Because predicted and ground-truth parts do not share a common parameterization, labels are transferred via nearest-neighbor matching from predicted surfaces to points sampled on the ground-truth mesh.
Let $\mathcal{X}{=}\{\mathbf{x}_n\}_{n=1}^{N}$ denote the sampled surface points, and let
$\mathcal{G}{=}\{G_i\}_{i=1}^{K_g}$ and $\mathcal{P}{=}\{P_j\}_{j=1}^{K_p}$ denote the ground-truth and predicted partitions over $\mathcal{X}$, where $G_i,P_j \subseteq \mathcal{X}$.
For two regions $A,B \subseteq \mathcal{X}$, their intersection-over-union is defined as
\begin{equation*}
    \mathrm{IoU}(A,B)
    =
    \frac{|A \cap B|}{|A \cup B|}.
\end{equation*}
Following the class-agnostic part-segmentation protocol commonly used on
PartObjaverse-Tiny~\cite{sampart3d_yang2024}, we report mean IoU (mIoU) by greedily
matching every ground-truth part to its best-overlapping predicted part:
\begin{equation*}
    \mathrm{mIoU}
    =
    \frac{1}{K_g}
    \sum_{i=1}^{K_g}
    \max_{1\leq j\leq K_p} \mathrm{IoU}(G_i, P_j).
\end{equation*}
Because each ground-truth part picks its best predicted match independently, the
mapping is allowed to be many-to-one; this avoids the brittleness of optimal
(Hungarian) assignment under heavily over- or under-segmented predictions.

To assess structural correctness around articulations, we additionally compute Rand Index (RI), Variation of Information (VoI), and Segmentation Covering (SC).
Let $\mathcal{C}{=}\{(u,v)\mid 1\leq u<v\leq N\}$ be the set of all point pairs, and let
$\delta_{\mathcal{S}}(u,v){=}1$ if $\mathbf{x}_u$ and $\mathbf{x}_v$ belong to the same segment under partition $\mathcal{S}$, and $0$ otherwise.
The Rand Index is defined as
\begin{equation*}
    \mathrm{RI}
    =
    \frac{1}{|\mathcal{C}|}
    \sum_{(u,v)\in\mathcal{C}}
    \mathbf{1}
    \!\left[
    \delta_{\mathcal{G}}(u,v)
    =
    \delta_{\mathcal{P}}(u,v)
    \right].
\end{equation*}
VoI is the information-theoretic distance between the two partitions:
\begin{equation*}
    \mathrm{VoI}
    =
    H(\mathcal{G}) + H(\mathcal{P})
    - 2\,I(\mathcal{G};\mathcal{P}),
\end{equation*}
where $H(\cdot)$ denotes partition entropy and $I(\cdot;\cdot)$ denotes mutual
information.
Segmentation Covering measures the average best predicted coverage of each
ground-truth part, weighted by its size:
\begin{equation*}
    \mathrm{SC}
    =
    \sum_{i=1}^{K_g}
    \frac{|G_i|}{N}
    \,\max_{1\leq j\leq K_p}
    \mathrm{IoU}(G_i,P_j).
\end{equation*}

We compare Hi-TOPS against primitive-based abstraction methods~\cite{primAny, mps_2023, ems_2022} (converted to surface partitions) and recent 3D part segmentation models~\cite{sampart3d_yang2024, s2am3d_su2025, parfield_2025, partsam_zhu2025, pointsam_zhou2025}.

\mypara{Setup}
We use a single parameter setting for \textsc{Hi-TOPS} across every dataset and every shape: the resolution set is fixed to $R{=}\{16,32,64\}$, the per-face barycentric sample count to $N_\mathrm{face}{=}36$, the four cue weights to $\boldsymbol{\lambda}{=}(1,1,1,1)$, the Freeze ratio to $\alpha{=}0.15$, and the dihedral edge-cut threshold to the $90$-th quantile of the per-mesh dihedral distribution ($q{=}90$). These five settings dominate the structural behaviour of the pipeline; all other constants are fixed defaults listed in Tab.~\ref{tab:hyperparams}
, and no per-shape or per-dataset tuning is applied.
All baselines are evaluated from their official GitHub repositories using their released implementations, configurations, and checkpoints whenever available. For primitive-based abstraction methods, their primitive outputs are converted to surface partitions before evaluation. Specifically, each sampled surface point is assigned to the nearest predicted primitive support or nearest predicted surface element, and all metrics are then computed on the same sampled point set $\mathcal{X}$. No manual correction is applied to the predicted labels.

\mypara{Hyperparameters}
Tab.~\ref{tab:hyperparams} lists the full hyperparameter set used across all experiments. Defaults remain fixed across PartObjaverse-Tiny, PartNet, and HY3D-Bench; per-shape tuning is never applied.

\begin{table}[t]
\centering
\small
\caption{Hyperparameter table for \textsc{Hi-TOPS}. All defaults are dataset-agnostic.}
\label{tab:hyperparams}
\begin{tabularx}{\linewidth}{@{}>{\raggedright\arraybackslash}X r@{}}
\toprule
Symbol \& meaning & Default \\
\midrule
\multicolumn{2}{@{}l}{\emph{Cues \& Topology Score (App.~\ref{app:cues})}}\\
$R$ --- resolution set                                                                & $\{16,32,64\}$ \\
$\boldsymbol{\lambda}$ --- cue weights (Eq.3)                                        & $(1.0,1.0,1.0,1.0)$ \\
$N_\mathrm{face}$ --- per-face barycentric samples                                              & 
$36$ \\
$\alpha$ --- Freeze ratio (Eq.4, 5; same cut drives both)               & $0.15$ \\
\midrule
\multicolumn{2}{@{}l}{\emph{SDF / SQ fitting (App.~\ref{app:sdf})}}\\
$\tau$ --- TSDF truncation                                                                      & $8\,\Delta^{(64)}$ \\
Flow-zone tol.\ --- mesh hull dilation                                                       & $2\,\Delta^{r_\mathrm{fit}}$ \\
$\{\eta\}$ --- a-scale retry trust-region multipliers                                         & $\{1.0,0.5,0.25\}$ \\
EDT depth threshold --- body $\to$ surface seed switch                                          & $1.5$ vox \\
$V_\mathrm{min}$ --- accept volume (body / thin)                                                & $5$ / $3$ vox \\
$\mu_\mathrm{ext}$ --- exterior intrusion penalty                                               & $5$ \\
\midrule
\multicolumn{2}{@{}l}{\emph{Curvature-aware edge primitives (App.~\ref{app:curv})}}\\
$q$ --- dihedral cut quantile                                                                   & $90$ \\

boundary-smoothing iterations                                                                   & $1$ \\
\bottomrule
\end{tabularx}
\end{table}

\section{Geometric Cue Details}
\label{app:cues}

This appendix details the five voxel-level cues entering the Topology Score of Eq.3: the within-block curvature mean $\mu_i^\kappa$, the within-block curvature standard deviation $\sigma_i^\kappa$, the cross-block curvature gradient $\Delta_i^\kappa$, the normal-inconsistency cue $\phi_{i,\mathrm{norm}}^{(r)}$, and the density factor $\phi_{i,\mathrm{den}}^{(r)}=n_i^{(r)}$. We first describe the per-sample primitives that feed the aggregation, then the single-pass block aggregation that produces the four bracketed terms of Eq.3 in left-to-right order, and finally the two second-order curvature statistics ($\sigma^\kappa$, $\Delta^\kappa$) and the density modulation. The sensitivity analysis of the four cue weights $\boldsymbol{\lambda}$ in Eq.3 is deferred to App.~\ref{app:cue_weight}.

\textbf{Surface sampling.} For each face $f\in\mathcal{F}$ we instantiate a deterministic \emph{uniform barycentric grid} of $N_\mathrm{face}$ points $(u,v,w)$ with $u{+}v{+}w{=}1$, and lift each grid point to $\mathbb{R}^3$ by $\mathbf{p}=u\mathbf{v}_0{+}v\mathbf{v}_1{+}w\mathbf{v}_2$. Sampling at the face level (rather than mesh-globally) guarantees that small triangles still contribute samples; the deterministic grid is preferred over stochastic alternatives so that the per-block sample counts---used as a confidence weight in $S_B^{(r)}$---are reproducible per shape.
 
\textbf{Per-sample mean curvature($\phi_\mathrm{curv}$).} We compute the per-vertex mean curvature $\kappa_v=\|\mathbf{H}_v\|/2$ from the standard cotangent Laplace--Beltrami operator with mixed-Voronoi vertex areas. We then clip $\{\kappa_v\}$ at the $99.5$-th percentile to suppress tessellation outliers, normalize by the global maximum, and broadcast to faces by $c_f=\tfrac{1}{3}\sum_{v\in f}\kappa_v\in[0,1]$. The same $c_f$ is shared by all $N_\mathrm{face}$ samples of $f$; equivalently, this defines the per-sample curvature primitive $\phi_\mathrm{curv}(\mathbf{p})\equiv c_{f(\mathbf{p})}$, whose within-block average is the term $\mu_i^\kappa$ in Eq.3.

\textbf{Per-sample normal deviation($\phi_\mathrm{norm}$).} For each face $f$ we retrieve its $k{=}8$ centroid-nearest faces $\mathcal{N}_f$ via a KD-tree and set
\[
\phi_{\mathrm{norm}}(f)=\tfrac{2}{\pi}\!\cdot\!\frac{1}{|\mathcal{N}_f|}\!\sum_{f'\in\mathcal{N}_f}\!\arccos\langle\mathbf{n}_f,\mathbf{n}_{f'}\rangle,
\]
which is likewise broadcast to all $N_\mathrm{face}$ samples of $f$. The $k{=}8$ neighbourhood balances locality against robustness to a single noisy adjacency. The $2/\pi$ factor expresses the mean deviation in units of a right angle, placing $\phi_{\mathrm{norm}}$ on the same order as the normalized curvature $c_f$; values above one occur only where neighbouring normals turn by more than $90^\circ$, as at thin sheets, and are kept rather than clipped since such regions are precisely those the Freeze class should protect.

\textbf{Block aggregation.} Each sample $\mathbf{p}\in[-0.5,0.5]^3$ is hashed into the $r^3$ grid by the index map $\iota_r(\mathbf{p})=\lfloor(\mathbf{p}+0.5)r\rfloor$, after which we run a single $r^3$-bucket accumulation over \emph{all} samples of \emph{all} faces. Denoting the sample set falling in voxel $v_i^{(r)}$ by $\mathcal{P}_i^{(r)}$ and its size by $n_i^{(r)}$, the four quantities entering Eq.3 are, in the same left-to-right order as the bracketed terms and the density factor of Eq.3,
\begin{align*}
\mu_i^\kappa &= \tfrac{1}{n_i^{(r)}}\!\!\sum_{\mathbf{p}\in\mathcal{P}_i^{(r)}}\!\!c_{f(\mathbf{p})}, \quad
\sigma_i^\kappa = \sqrt{\overline{c^2}_i-(\mu_i^\kappa)^2}, \notag\\
\phi_{i,\mathrm{norm}}^{(r)} &= \tfrac{1}{n_i^{(r)}}\!\!\sum_{\mathbf{p}\in\mathcal{P}_i^{(r)}}\!\!\phi_{\mathrm{norm}}(f(\mathbf{p})),\quad
\phi_{i,\mathrm{den}}^{(r)}=n_i^{(r)},
\label{eq:cue_aggregation}
\end{align*}
i.e.\ $\mu_i^\kappa\!\leftrightarrow\!\lambda_1$-term, $\sigma_i^\kappa\!\leftrightarrow\!\lambda_2$-term, $\phi_{i,\mathrm{norm}}^{(r)}\!\leftrightarrow\!\lambda_4$-term, and $\phi_{i,\mathrm{den}}^{(r)}$ enters the $\sqrt{\cdot}$ modulator. The remaining $\lambda_3$-term $\Delta_i^\kappa$ is not produced by this single-pass bucket and is derived afterwards from the dense $\mu^\kappa$ field, as detailed below. Block means are computed with a masked reciprocal ($1/n^{(r)}_i$ evaluated only where $n^{(r)}_i>0$); voxels with no sample support are excluded from the score field altogether.

\textbf{Within-block curvature spread ($\sigma_i^\kappa$).} The block mean $\mu_i^\kappa$ alone dilutes seams that share a block with smooth area, so we additionally summarize each block by its standard deviation $\sigma_i^\kappa=\sqrt{\overline{c^2}_i-(\mu_i^\kappa)^2}$, where $\overline{c^2}_i$ is the second-moment accumulator collected in the same single bucket loop as $\mu_i^\kappa$ (variance identity $\mathrm{Var}(c)=\mathbb{E}[c^2]-(\mathbb{E}[c])^2$, no second pass; the radicand is clipped at zero to absorb floating-point negatives in nearly homogeneous blocks). $\sigma^\kappa$ fires whenever low- and high-curvature samples coexist in the same voxel, which is the dominant signature of joints where a flat support meets a curved articulation inside a single block. The std (rather than the variance) is chosen so that $\sigma^\kappa$ shares the same dimension as $\mu^\kappa$ and $\Delta^\kappa$, which is what allows the three curvature terms in Eq.3 to be summed under a common weight scale.

\textbf{Cross-block curvature gradient ($\Delta_i^\kappa$).} Within-block statistics miss ridges aligned with block boundaries, which leave a step in the dense $\mu^\kappa$ field rather than a within-block spread. We capture such transitions by $\Delta_i^\kappa=\|\nabla\mu^\kappa\|_2(v_i^{(r)})$, where the components $(\partial_x,\partial_y,\partial_z)\mu^\kappa$ are obtained on the dense $r^3$ array $\{\mu_i^\kappa\}$ (second-order central differences inside, first-order one-sided at the grid boundary), combined into a per-voxel $\ell_2$ magnitude.
The differences are taken with unit voxel spacing. Thus, $\Delta_i^\kappa$ measures per-block curvature contrast rather than a spacing-normalized metric gradient. This keeps it comparable to the normalized cues $\mu^\kappa$ and $\sigma^\kappa$; dividing by $d_v^{(r)}=1/r$ would introduce an artificial $O(r)$ scale factor and make the cue balance drift across resolutions. Therefore, $\Delta^\kappa$ complements $\sigma^\kappa$: $\sigma^\kappa$ measures heterogeneity \emph{within} a block, $\Delta^\kappa$ measures heterogeneity \emph{between} a block and its neighbours.

\textbf{Density modulation.} The factor $\sqrt{\phi_{i,\mathrm{den}}^{(r)}}=\sqrt{n_i^{(r)}}$ in Eq.3 acts as a confidence-weighted multiplier: blocks with more sample support receive proportionally larger scores, suppressing spurious responses from sparsely sampled fragments. The square root is preferred over a linear $n_i^{(r)}$, which would over-reward densely tessellated patches: because each face contributes a fixed number of samples, subdividing the same surface patch increases $n_i^{(r)}$ without adding geometric evidence. The square root compresses this tessellation-induced growth, whereas $\log(\cdot)$ would saturate too early. 

\section{Analysis with PartSAM and PartField}
\label{app:ana_part}
Table 1. reports mIoU on the three benchmarks together with average inference time; detailed statistics of the three evaluation subsets are summarized in Tables~\ref{tab:dataset_overview}--\ref{tab:partobjaverse_instance_statistics} of this appendix. The three datasets pose distinctly different challenges, which directly explain the complementary behavior we observe across methods.
\paragraph{Dataset difficulty profile.} As shown in Table~\ref{tab:dataset_overview}, the three subsets occupy different regions of the part-count spectrum. PartObjaverse-Tiny has the lowest complexity (mean $10.56$, max $25$, std $4.52$) and serves as a moderate-granularity, open-vocabulary instance segmentation benchmark spanning eight semantic super-categories. HY3D-Bench is heavier (mean $14.80$, max $50$) and strongly long-tailed: $25.5\%$ of its objects contain $\geq 21$ parts, yet it provides no category labels, making it a multi-part Objaverse-style distribution test. PartNet selected, although averaging $15.34$ leaves, exhibits by far the most extreme tail (max $96$, std $14.80$). Table~\ref{tab:partnet_leaf_part_counts} further shows that this tail is concentrated in \textit{StorageFurniture} (mean $26.50$, max $96$) and \textit{Bed} (mean $22.65$, max $89$), reflecting the fine-grained CAD-style decomposition (drawers, slats, frame elements) that PartNet supplies.
\paragraph{PartSAM} PartSAM is the fastest method ($20.05$s) and achieves the best result on HY3D-Bench ($55.13$), where its learned semantic priors generalize well to Objaverse-style open-domain objects. However, it collapses on PartNet ($30.16$), where the dominant difficulty comes precisely from the highly fragmented \textit{StorageFurniture} and \textit{Bed} categories visible in Table~\ref{tab:partnet_leaf_part_counts}: PartSAM recovers semantically reasonable coarse regions but lacks the capacity to separate small repetitive parts such as individual drawers, slats, or frame components. In other words, its learned prior is biased toward semantically prominent regions and underperforms when the ground truth requires fine geometric structure.

\paragraph{PartField} PartField is strongest on PartObjaverse-Tiny ($66.77$), with a clear margin over both baselines. Table~\ref{tab:partobjaverse_instance_statistics} indicates why: the dominant categories are \textit{Transportations} ($19.0\%$, mean $14.53$ parts), \textit{Electronics} ($17.0\%$, $9.18$), and \textit{Human-Shape} ($14.5\%$, $12.17$)---all open-category objects of moderate per-instance complexity that align well with the distribution under which PartField's learned field representation was trained. Once the distribution shifts to fine-grained CAD leaves (PartNet, $45.57$) or to long-tailed unlabeled Objaverse-style meshes (HY3D-Bench, $44.56$), the learned prior no longer transfers, and the score drops by roughly $20$ mIoU.
\paragraph{Hi-TOPS.} Hi-TOPS is the only training-free method, yet it obtains the best result on PartNet ($55.87$, $+10.3$ over PartField and $+25.7$ over PartSAM) and remains a close second on the other two benchmarks ($51.63$ on PartObjaverse-Tiny, $47.22$ on HY3D-Bench). Crucially, its cross-dataset mIoU standard deviation is only $3.5$, compared to $\approx 10.5$ for both trained baselines---a direct consequence of relying on an explicit meso-scale topology prior rather than a learned distribution. This consistency is especially valuable on PartNet, where Table~\ref{tab:partnet_leaf_part_counts} shows that $19$ of the $100$ objects fall into the $21$--$50$ bucket and $3$ exceed $50$ leaves; learned methods either over-merge (PartSAM, $30.16$) or partially over-merge (PartField, $45.57$) on these extreme cases, whereas the structural prior used by Hi-TOPS scales naturally with the true count. The main cost is runtime: at $129.5$\,s per object Hi-TOPS is $\sim 2\times$ slower than PartField and $\sim 6.6\times$ slower than PartSAM, due to the explicit superquadric fitting stage.

\paragraph{Summary.} The three methods therefore occupy clearly distinct operating points that are well-aligned with the dataset profile in Table~\ref{tab:dataset_overview}: PartSAM is suited to fast bulk processing of open-domain meshes with moderate part counts, PartField excels on open
-vocabulary categories with distributions close to its training data, and Hi-TOPS provides the highest-quality, distribution-agnostic decomposition---particularly on extreme long-tailed structural cases revealed by PartNet's statistics.

\begin{table}[t]
    \centering
    \caption{Overview of the three evaluation subsets. PartNet is counted by the number of instance part; PartObjaverse-Tiny uses instance-level labels; HY3D-Bench uses the number of part mesh.}
    \label{tab:dataset_overview}
    \resizebox{\linewidth}{!}{
    \begin{tabular}{lcccccc}
        \toprule
        Dataset & N  & Mean & Median & Min & Max & Std \\
        \midrule
        PartNet selected & 100  & \textbf{15.34} & 11.5 & 2 & \textbf{96} & 14.80 \\
        PartObjaverse-Tiny & 200  & 10.56 & 10 & 3 & 25 & 4.52 \\
        HY3D-Bench & 200 & 14.80 & 12 & 2 & 50 & 11.23 \\
        \bottomrule
    \end{tabular}}
\end{table}

\begin{table}[t]
    \centering
    \caption{Category distribution and part counts of PartNet selected.}
    \label{tab:partnet_leaf_part_counts}
    \resizebox{\linewidth}{!}{
    \begin{tabular}{lcccccc}
        \toprule
        Category & N & Mean & Median & Min & Max & Std \\
        \midrule
        Bed & 20 & 22.65 & 19.5 & 4 & 89 & 18.36 \\
        Chair & 20 & 11.20 & 11 & 4 & 24 & 4.51 \\
        Lamp & 20 & 4.85 & 4 & 2 & 10 & 1.90 \\
        StorageFurniture & 20 & \textbf{26.50} & 21.5 & 10 & \textbf{96} & \textbf{19.45} \\
        Table & 20 & 11.50 & 10 & 3 & 28 & 6.03 \\
        \midrule
        \textbf{Overall} & \textbf{100} & \textbf{15.34} & \textbf{11.5} & \textbf{2} & \textbf{96} & \textbf{14.80} \\
        \bottomrule
    \end{tabular}}
\end{table}

\begin{table}[t]
    \centering
    \caption{Category distribution and instance part-count statistics of PartObjaverse-Tiny.}
    \label{tab:partobjaverse_instance_statistics}
    \resizebox{\linewidth}{!}{
    \begin{tabular}{lccccc}
        \toprule
        Category & N & Percentage & Mean & Median & Max \\
        \midrule
        Transportations & 38 & 19.0\% & \textbf{14.53} & 13 & \textbf{25} \\
        Electronics & 34 & 17.0\% & 9.18 & 9 & 17 \\
        Human-Shape & 29 & 14.5\% & 12.17 & 13 & 16 \\
        Daily-Used & 25 & 12.5\% & 6.60 & 6 & 12 \\
        Buildings\&Outdoor & 25 & 12.5\% & 10.36 & 10 & 18 \\
        Animals & 23 & 11.5\% & 12.04 & 12 & 21 \\
        Plants & 18 & 9.0\% & 8.44 & 8 & 23 \\
        Food & 8 & 4.0\% & 5.25 & 4.5 & 10 \\
        \midrule
        \textbf{Total} & \textbf{200} & \textbf{100\%} & \textbf{10.56} & \textbf{10} & \textbf{25} \\
        \bottomrule
    \end{tabular}}
\end{table}

\section{Additional Experiments}
\label{app:add_ablation}
\subsection{Ablation Protocol}
Table 3 and A.5 use a fixed diagnostic fitting branch to isolate cue and threshold effects, whereas Table 4 and A.6 use the final Hi-TOPS  fitting pipeline. Their absolute metric values are therefore not directly comparable across the two groups. All ablations use a fixed 36-shape PartObjaverse-Tiny subset;
\subsection{Search for Cues' Weights}
\label{app:cue_weight}
The topology score is computed from three voxel-level cues derived from surface samples—curvature (estimated via the cotangent Laplacian and aggregated by mean, standard deviation, and gradient magnitude), local normal variation, and surface sample density (uniformly sample 36 points on each face). Cue weights were determined via grid search. We found that uniform setting (1,1,1,1) yields stable performance across scenes and therefore adopt it as the default configuration. The table has been reorganized for clarity (same setting in Table 3). The score is intended as a simple and robust proxy rather than a fully optimized metric.
\begin{table}[h]
    \centering
    \caption{Sensitivity analysis of cue weights.}
    \label{tab:cue_weight_sensitivity}
    \scalebox{0.95}{
    \begin{tabular}{ccccc}
        \toprule
        Cue Weights & RI $\uparrow$ & VoI $\downarrow$ & SC $\uparrow$ & mIoU $\uparrow$ \\
        \midrule
        $(1,0.5,1,1)$   & 0.704 & 1.937 & 0.344 & 32.01 \\
        $(1,2,1,1)$     & \textbf{0.723} & 1.864 & 0.380 & 33.09 \\
        $(1,1,0.5,1)$   & 0.718 & 1.855 & 0.385 & 32.72 \\
        $(1,1,2,1)$     & 0.709 & 1.878 & 0.361 & 32.50 \\
        $(1,1,1,0.5)$   & 0.721 & \textbf{1.780} & 0.390 & 34.10 \\
        $(1,1,1,2)$     & 0.712 & 1.788 & 0.390 & 34.01 \\
        $(1,1,1,1)$     & 0.715 & 1.820 & \textbf{0.394} & \textbf{34.67} \\
        \bottomrule
    \end{tabular}}
\end{table}

\subsection{Runtime Decomposition}
\label{app:runtime}
All runtime results are measured on a single CPU thread using an AMD Ryzen 9 5900X processor. 
Table~\ref{tab:resolution_runtime_breakdown} shows that topology scoring is relatively inexpensive, while SQ fitting dominates the total runtime and grows rapidly with resolution. 
The full pipeline increases from 38.06s at $16^3$ to 101.63s at $64^3$ and 234.07s at $128^3$. 
This confirms that finer voxel resolutions incur a substantially higher fitting cost without yielding better performance, motivating our hierarchical-resolution design.

\textbf{Cost of removing TSDF guidance.} Table~4 shows that dropping the TSDF term \emph{increases} runtime at every resolution (101.6\,s vs.\ 60.4\,s at $r{=}64$; 238.4\,s vs.\ 132.1\,s for the full hierarchy). Without the smooth TSDF residual the fitter has only discrete voxel evidence to align to, so each solve consumes more shrink-and-retry rounds and each accepted primitive claims less volume, requiring more seeds before the coverage criterion of Alg.~A.2 is met. TSDF guidance therefore improves accuracy and cost simultaneously.

\begin{table}[t]
    \centering
    \caption{Resolution sensitivity and CPU runtime breakdown on a subset of PartObjaverse-Tiny.}
    \label{tab:resolution_runtime_breakdown}
    \resizebox{\linewidth}{!}{
    \begin{tabular}{ccccccccc}
        \toprule
        Res. & RI $\uparrow$ & VoI $\downarrow$ & SC $\uparrow$ & mIoU $\uparrow$ 
        & Score & SQ & Seg. & Total \\
        & & & & & \multicolumn{4}{c}{CPU time (s)} \\
        \midrule
        16  & 0.716 & 2.132 & 0.348 & 30.68 & 7.89  & 18.7 & 11.47 & 38.06 \\
        32  & 0.715 & 1.820 & 0.394 & 34.67 & 7.83  & 24.64  & 12.03 & 44.50 \\
        64  & 0.672 & 1.774 & 0.386 & 32.52 & 8.23  & 75.69 & 17.71 & 101.63 \\
        128 & 0.675 & 1.618 & 0.412 & 31.06 & 19.72 & 176.4 & 37.95 & 234.07 \\
        \bottomrule
    \end{tabular}}
\end{table}




    
        

\section{Full Algorithm Overview}
\label{app:algo}

We provide a self-contained pseudocode of \textsc{Hi-TOPS} that covers normalization, per-resolution cue extraction, Topology Score, hierarchical carrier construction, TSDF supervision, coarse-to-fine SDF-constrained SQ fitting (with Flow-zone retry), and curvature-aware Surface Assignment. Cross-references to main-text equations and subsequent appendices are anchored on each line. Symbols introduced without definition are listed in Tab.~\ref{tab:hyperparams}.


\subsection{Rule-based Freeze Fusion}
\label{app:adap}\label{app:hier}

This appendix formalizes the four fusion rules of Sec.3.2 together with the two housekeeping mechanisms (sparse-support refinement and backward consistency) that the implementation applies on top of them.

\textbf{Input.} At each $r\in\{16,32,64\}$, the topology score $S_B^{(r)}$ and the top-$\alpha$ cut of Eq.4, 5  yield a per-voxel binary class $c_i^{(r)}\in\{\textsc{Flow},\textsc{Freeze}\}$ on the surface support $\Omega^{(r)}$. The fusion below consumes the three binary maps and returns a single non-overlapping octree carrier whose leaves cover $\Omega^{(64)}$.

\textbf{Octree topology.} The $r{=}16$ grid forms the root layer ($16^3$ candidate roots); each non-empty root has $8$ children at $r{=}32$, and each non-empty $r{=}32$ child has $8$ grandchildren at $r{=}64$. Empty roots and empty children (no surface support) are trivial leaves and never split.

\textbf{Notation.} For a parent block at resolution $r\in\{16,32\}$ with class $c_p$, let $\{c_{c_1},\dots,c_{c_8}\}$ denote the classes of its eight finer children at resolution $2r$, $\mathcal{N}=\{i:c_{c_i}\text{ has surface support}\}$ the set of non-empty children with $n_\mathrm{nz}=|\mathcal{N}|$, and $n_\mathrm{diff}=|\{i\in\mathcal{N}:c_{c_i}\neq c_p\}|$ the number of non-empty children that disagree with the parent.

\textbf{Four fusion rules.} The parent decides between staying as a leaf and refining into its eight children according to the four mutually exclusive configurations below; when the parent stays leaf, all of its descendants inherit $c_p$; when the parent refines, each child carries its own class $c_{c_i}$.
\begin{description}
  \item[\textbf{(i) Coarse-flow preservation.}] If $c_p=\textsc{Flow}$ and $\forall\,i\in\mathcal{N}:c_{c_i}=\textsc{Flow}$, the parent is kept as a single Flow leaf, retaining stable body-level support.
  \item[\textbf{(ii) Fine-freeze promotion.}] If $c_p=\textsc{Flow}$ and $\exists\,i\in\mathcal{N}:c_{c_i}=\textsc{Freeze}$, the parent is refined. A single emerging Freeze child is already sufficient: any sub-region that has crossed the Freeze threshold at the finer scale --- whether a thin joint or a local articulation --- is preserved instead of being smoothed away by the coarse Flow label.
  \item[\textbf{(iii) Coarse-freeze inheritance.}] If $c_p=\textsc{Freeze}$ and $n_\mathrm{diff}/n_\mathrm{nz}<0.5$ (the majority of non-empty children remain Freeze), the parent is kept as a single Freeze leaf, maintaining stable large-scale separations along sharp transitions.
  \item[\textbf{(iv) Coarse-freeze relaxation.}] If $c_p=\textsc{Freeze}$ and $n_\mathrm{diff}/n_\mathrm{nz}\ge 0.5$ (the majority of non-empty children fall back to Flow), the parent is refined so that the finer-scale evidence overrides the coarse Freeze decision and the still-Freeze children remain as Freeze leaves at the finer resolution $2r$.
\end{description}

\textbf{Sparse-support refinement.} Independently of the four rules, any non-empty parent with $n_\mathrm{nz}\le 2$ is refined: a class supported by only one or two finer children is statistically unreliable, so the block defers its class to the finer scale rather than committing to a coarse-leaf decision.

\textbf{Backward consistency.} The two octree levels are processed independently in the same pass: the $16{\to}32$ decision uses the $r{=}16$ parent and its eight $r{=}32$ children, and each $32{\to}64$ decision uses one $r{=}32$ child and its eight $r{=}64$ grandchildren. To keep the carrier a non-overlapping partition, if a $r{=}16$ parent does not split but at least one of its $r{=}32$ children splits to $r{=}64$, the $r{=}16$ parent is forced to split; the $r{=}32$ children that themselves did not split then inherit the $r{=}16$ parent's class. This preserves the invariant that no $r{=}16$ leaf shadows a $r{=}64$ leaf in the dense projection.

\textbf{Output projection.} The fused leaves are dense-projected onto a $64^3$ grid by assigning each $r{=}64$ voxel in the surface support the class of its enclosing leaf, yielding the projected priors $(\mathcal{B}_\mathrm{flow}^\mathrm{hier},\mathcal{B}_\mathrm{freeze}^\mathrm{hier})$ that drive the SDF-guided SQ fitter of App.~\ref{app:sdf}.


\subsection{Body-Surface SuperQuadric Fitting Details}
\label{app:sdf}

This appendix supplements the Sec.3.3: (A.1)--(A.6) elaborate \emph{Body-Surface SQ Initialization}, and (B.1)--(B.3) elaborate \emph{TSDF-Guided SQ Inflation}, i.e.\ the discrete realization of Eq.7 and the $\beta$-step accept/reject of Eq.~8. The mesh TSDF $d_{\mathcal{M}}$ is precomputed on the same $r{=}64$ grid as $\mathcal{G}_\mathrm{hier}$, clipped to $\pm\tau$ with $\tau=8\Delta^{(64)}$.

\textbf{(A.1) Volumetric constraints $\Omega_\mathrm{flow}$ and $\Omega_\mathrm{freeze}$.} The projected priors $(\mathcal{B}_\mathrm{flow}^\mathrm{hier},\mathcal{B}_\mathrm{freeze}^\mathrm{hier})$ delivered by App.~\ref{app:hier} are upgraded into two complementary voxel-space zones used as hard constraints by the SQ fitter. The \emph{Freeze zone} $\Omega_\mathrm{freeze}=\mathcal{B}_\mathrm{freeze}^\mathrm{hier}\cup\mathrm{ext\_void}$ collects all voxels the fitter must avoid, where $\mathrm{ext\_void}=\neg\,\mathrm{fillholes}(\mathcal{B}_\mathrm{flow}^\mathrm{hier}\cup\mathcal{B}_\mathrm{freeze}^\mathrm{hier})$ is the mesh exterior obtained by hole-filling the carrier and complementing; symmetrically, the \emph{Flow zone} $\Omega_\mathrm{flow}=\mathrm{dilate}(\neg\,\mathrm{ext\_void},\,2\Delta^{r_\mathrm{fit}})$ dilates the mesh hull outward by two voxels at the fit resolution to absorb the unavoidable overshoot of a smooth implicit SQ across a curved boundary. A Euclidean distance transform $\mathrm{EDT}_{\Omega_\mathrm{flow}}$ is precomputed once to cap the initial SQ radius at each seed. Together $\Omega_\mathrm{flow}$ and $\Omega_\mathrm{freeze}$ are the concrete realization of ``SQs grow inside Flow regions and are constrained near Freeze regions'' --- the carrier-level binary labels $\mathcal{B}_\mathrm{flow/freeze}^\mathrm{hier}$ are thus upgraded into geometric zones with explicit 2-voxel overshoot tolerance and explicit external-void exclusion.

\textbf{(A.2) Body seed: EDT maxima.} Body seeds are drawn from the available Flow body $\Omega_\mathrm{Body}=\mathcal{B}_\mathrm{flow}^\mathrm{hier}\cap\Omega_\mathrm{valid}\setminus(\Omega_\mathrm{claimed}\cup\Omega_\mathrm{freeze})$ by taking the global maximum of an EDT on $\Omega_\mathrm{Body}$,\[
\mathbf{s}=\arg\max_{\mathbf{v}\in\Omega_\mathrm{Body}}\mathrm{EDT}(\mathbf{v}).
\] Geometrically, this EDT depth equals the radius of the largest inscribed ball at $\mathbf{v}$, so its maxima coincide with the centers of locally thickest body cross-sections --- the ``dominant volumetric interiors'' described in the main text. After a fit is accepted and $\Omega_\mathrm{claimed}$ grows, the EDT on $\Omega_\mathrm{Body}$ is recomputed incrementally rather than from scratch, restricting work to voxels whose nearest forbidden voxel has changed.

\textbf{(A.3) Surface seed: residual-component centroids.} When the Body strategy is exhausted, the remaining $\Omega_\mathrm{Body}$ consists of thin attachments left untouched by EDT-based seeding. Surface seeds are placed at the geometric centroid of the largest unprocessed 6-connected component of $\Omega_\mathrm{Body}$. The centroid lies along the axis of a thin attachment rather than at its tip, providing the elongated TSDF window required for a meso-scale SQ to grow along the part axis.

\textbf{(A.4) Hierarchical Body--Surface switch.} The two strategies share a single scheduler \textsc{BodyOrSurfaceSeed}: at each iteration, if $\max_{\mathbf{v}\in\Omega_\mathrm{Body}}\mathrm{EDT}(\mathbf{v})\ge 1.5\Delta^{r_\mathrm{fit}}$ a Body seed is emitted, otherwise the call falls back to a Surface seed. The threshold is chosen so that a fit can still admit at least a single-voxel inscribed ball under the Flow-zone cap; below it, no body-style SQ can be born safely. Because the test is per-seed rather than per-stage, Body and Surface seeds may interleave inside a single $r$-stage, matching the main-text statement that the two strategies are not tied to a single resolution.

\textbf{(A.5) InitSQ.} Each seed $\mathbf{s}$ initializes $\boldsymbol{\theta}_0$ with center $\mathbf{c}_0=\mathbf{s}$, axis-aligned orientation (no PCA on the local ROI), equal semi-axes $a_x{=}a_y{=}a_z=\min(\rho_0,\mathrm{EDT}_{\Omega_\mathrm{flow}}(\mathbf{s}))$ with $\rho_0\!\propto\!\Delta^{r_\mathrm{fit}}$, and shape exponents $\epsilon_1{=}\epsilon_2{=}1$ (an ellipsoid). The optimization bounds $[\mathrm{lb},\mathrm{ub}]$ are derived from $\mathbf{s}$ and $r_\mathrm{fit}$ (Alg.~\ref{alg:sdf_sq_fit}, line~14).

\textbf{(A.6) Coarse-to-fine schedule over $r\in\{16,32,64\}$.} Three stages indexed by $r$ progressively expand the territory available to the fitter, $\Omega_\mathrm{valid}=\{\mathcal{R}_\mathrm{map}\!\le\!r\}\setminus\Omega_\mathrm{freeze}$, while the reserved set $\Omega_\mathrm{spec}=\{\mathcal{R}_\mathrm{map}\!>\!r\}$ protects finer carriers from being claimed early. The fitting scale $r_\mathrm{fit}=\min(r,32)$ controls the geometric scale of the SQ itself (initial semi-axis $\Delta^{r_\mathrm{fit}}$, ROI radius, Flow-zone tolerance); the cap $r_\mathrm{fit}\!\le\!32$ prevents the $r{=}64$ stage from degenerating into a single-voxel initial ball. Combined with (A.2)--(A.4), the three stages naturally run Body-dominated at coarse $r$ and Surface-dominated at fine $r$, which is what the main text means by ``first establishes robust body cores and then supplements missing surface details''.

\textbf{(B.1) Shell samples and TSDF residual (Eq.7).} The shell $\mathcal{P}_\mathrm{shell}^t$ of Eq.7 is realized by the near-surface band $\mathcal{P}_\mathbf{s}=\{\mathbf{p}\in\mathrm{ROI}(\mathbf{s}):d_{\mathcal{M}}(\mathbf{p})<1.5\Delta\}$, with $\mathrm{ROI}(\mathbf{s})=\mathrm{ball}(\mathbf{s},\rho(r_\mathrm{fit}))\cap\Omega_\mathrm{valid}\setminus(\Omega_\mathrm{claimed}\cup\Omega_\mathrm{freeze})$; the band width $1.5\Delta$ plays the role of $d_\mathrm{step}$. The residual entering Eq.7 is $r_{\boldsymbol{\theta}}(\mathbf{p})=\tilde d_{\boldsymbol{\theta}}(\mathbf{p})-d_{\mathcal{M}}(\mathbf{p})$, two-stage truncated to $\pm 0.3\tau$ on near-surface samples and $\pm\tau$ elsewhere.

\textbf{(B.2) Region-aware weighting $w(\mathbf{p})$.} The weight $w(\mathbf{p})$ in Eq.7 splits the active set $\mathcal{P}_\mathrm{a}=\{\mathbf{p}\in\mathcal{P}_\mathbf{s}:|\tilde d_{\boldsymbol{\theta}_0}(\mathbf{p})|<3\tau\}$ into three regimes, each targeting a distinct failure mode of an unweighted least-squares fit:
\begin{itemize}
  \item \textbf{Exterior} ($d_{\mathcal{M}}(\mathbf{p})>0$): $w=1+\mu_\mathrm{ext}\max(0,-\tilde d_{\boldsymbol{\theta}_0}(\mathbf{p}))/\tau$ with $\mu_\mathrm{ext}=5$. The numerator is the SQ's intrusion depth at an exterior point (positive only when the SQ wrongly engulfs that point); the $1/\tau$ normalization saturates $w$ at $1+\mu_\mathrm{ext}=6$ when intrusion reaches one full truncation band.
  \item \textbf{Near-surface} ($|d_{\mathcal{M}}(\mathbf{p})|<0.3\tau$): $w=1$ together with the tighter $\pm 0.3\tau$ residual clip of (B.1), giving sharper boundary-aligning gradients without letting far-from-surface points dominate.
  \item \textbf{Interior} ($d_{\mathcal{M}}(\mathbf{p})<0$): a Gaussian inlier weight $w_\mathrm{Gauss}(\mathbf{p})$ inherited from MPS~\cite{mps_2023}, which down-weights deep-interior points distant from the SQ surface, preventing the SQ from over-expanding to ``reach'' points that should belong to other primitives.
\end{itemize}

\textbf{(B.3) Iterative update, acceptance, and termination (Eq.~8).} The $\beta\Delta\boldsymbol{\theta}^t$ update of Eq.~8 is realized by a TRF (trust-region reflective) bound-constrained least-squares solve
\[
\boldsymbol{\theta}^{*}=\arg\min_{\boldsymbol{\theta}\in[\mathrm{lb},\mathrm{ub}]}\sum_{\mathbf{p}\in\mathcal{P}_\mathrm{a}}w(\mathbf{p})\,r_{\boldsymbol{\theta}}(\mathbf{p})^{2},
\]
wrapped in a retry loop with $a_\mathrm{scale}{=}\eta\in\{1.0,0.5,0.25\}$ that halves the trust-region step along the semi-axis directions on each retry (rotation and translation steps unchanged); $\eta$ thus plays the role of the inflation step size $\beta$. After each solve we test the Flow-zone acceptance constraint
\[
\partial\Omega(\boldsymbol{\theta}^{*})\subseteq\Omega_\mathrm{flow}
\]
by sampling the SQ surface; any violation triggers shrink-and-retry. A fit that passes the test produces a settle mask and a net claim
\[
M_\mathrm{fit}=\{\mathbf{v}:\tilde d_{\boldsymbol{\theta}^{*}}(\mathbf{v})\le 2\Delta\},\qquad
\Delta\Omega=M_\mathrm{fit}\setminus(\Omega_\mathrm{freeze}\cup\Omega_\mathrm{claimed}\cup\Omega_\mathrm{spec}),
\]
and is accepted iff
\[
|\Delta\Omega|\ge V_\mathrm{min},
\]
with $V_\mathrm{min}{=}5$ voxels for body seeds and $3$ for surface seeds. Together these realize the main text's ``stays mostly within $\mathcal{B}_\mathrm{flow}$ without intruding into $\mathcal{B}_\mathrm{freeze}$'' acceptance. All three retries failing the Flow-zone test, or the final fit failing the volume threshold, lead to seed rejection (the iteration-budget / negligible-improvement case in Eq.~8); the seed location is then excluded from later iterations of the same stage.
\subsection{Curvature-Aware Mesh Over-segmentation}
\label{app:curv}

The SQ-to-Mesh Assignment step (Algorithm~\ref{alg:tops}, line~7) takes as input an over-segmentation $\{\mathcal{E}_j\}_{j=1}^{A}$ of $\mathcal{M}$ ($A{\sim}200$--$500$ edge primitives per shape), obtained by an edge-score pipeline on the face-adjacency graph of $\mathcal{M}$. Inspired by DORA's Sharp Edge Sampling~\cite{chen2025dora}, which detects salient edges as face pairs whose normals deviate beyond a threshold, we adopt the same dihedral-angle signal but repurpose it from a \emph{sampling} cue (selecting points along sharp edges) into a \emph{boundary-construction} cue (cutting the face graph along sharp edges so that connected components become edge-primitive interiors).

\textbf{Edge cut.} For each adjacency pair $(f_1,f_2)$ we score the shared edge by its dihedral angle, $s(f_1,f_2)=\theta_{f_1f_2}^{\circ}=\arccos\langle\mathbf{n}_{f_1},\mathbf{n}_{f_2}\rangle$, and sever the adjacency whenever $s$ exceeds the $q$-th quantile of $\{s\}$ over the mesh, with $q=90$ in all experiments; this quantile-based criterion adapts to each shape's overall sharpness distribution and keeps the resulting primitive count stable across datasets without per-shape tuning. After cutting, a flood-fill pass cleans up small fragments and smooths the primitive boundaries (one smoothing iteration, see Tab.~\ref{tab:hyperparams}), producing a per-face label vector $\boldsymbol{\ell}^{\mathrm{ep}}\!\in\!\{0,\dots,A{-}1\}^{|\mathcal{F}|}$ that feeds Eq.9.

\subsection{Surface Assignment}
\label{app:assign}

\textbf{Edge-primitive-level voting (Eq.9).} For each face $f$ we run a single nearest-neighbour query on the global voxel KD-tree $\mathcal{T}=\bigcup_k\mathcal{B}_k$ over all SQ-claimed voxel sets, obtaining $(\hat k_f,d_f)$. Within each edge primitive $\mathcal{E}_j$ the SQ label is determined by distance-weighted majority voting,
\[
k^*_j=\arg\max_k\sum_{f\in\mathcal{E}_j}\!\exp(-d_f/\tau_v)\,\mathds{1}[\hat k_f=k],
\]
so that the seam strictly follows the curvature-aware primitive boundary rather than the noisy per-face NN frontier. We use $\tau_v=\Delta^{(64)}$ as the default temperature, which makes a face whose NN distance is one voxel-edge contribute $1/e$ as much as a face exactly on the SQ surface.

\begin{algorithm}[!t]
\caption{\textsc{Hi-TOPS}: Topology-Scored Structural Prior for 3D Part Decomposition}
\label{alg:tops}
\begin{algorithmic}[1]
\Require Triangle mesh $\mathcal{M}$ normalized to $[-0.5,0.5]^3$; resolutions $R{=}\{16,32,64\}$; Freeze ratio $\alpha$
\Ensure Topology-consistent surface partition $\mathcal{S}{=}\{\mathcal{M}_k\}_{k=1}^{K}$

\For{each resolution $r\in R$}
    \State Voxelize $\mathcal{M}$, aggregate per-block geometric cues
    \hfill\Comment{App.~\ref{app:cues}}
    \State Compute Topology Score $S_B^{(r)}$, then split into Flow/Freeze by top-$\alpha$ cut
    \hfill\Comment{Eq.3, 4, 5}
\EndFor
\State Build hierarchical carrier $\mathcal{G}_{\mathrm{hier}}$ from $\{\mathcal{B}_\mathrm{flow}^{(r)},\mathcal{B}_\mathrm{freeze}^{(r)}\}$ and project to a $64^3$ grid
\hfill\Comment{App.~\ref{app:hier}}
\State Compute the truncated mesh SDF $d_{\mathcal{M}}$ on the carrier
\State $\mathcal{Q}\leftarrow\textsc{SDFSQFit}(\mathcal{G}_{\mathrm{hier}},d_{\mathcal{M}})$
\hfill\Comment{Alg.~\ref{alg:sdf_sq_fit}}
\State Over-segment $\mathcal{M}$ into curvature-aware edge primitives $\{\mathcal{E}_j\}$
\hfill\Comment{App.~\ref{app:curv}}
\State Assign each edge primitive to a primitive in $\mathcal{Q}$ by distance-weighted face voting
\hfill\Comment{Eq.9, App.~\ref{app:assign}}
\State \Return $\mathcal{S}=\bigl\{\bigcup_{j:k^*_j=k}\mathcal{E}_j\bigr\}_{k}$
\end{algorithmic}
\end{algorithm}

\begin{algorithm}[!t]
\caption{\textsc{SDFSQFit}: SDF-constrained Superquadric Fitting}
\label{alg:sdf_sq_fit}
\begin{algorithmic}[1]
\Require Hierarchical carrier $\mathcal{G}_{\mathrm{hier}}$ with class map and per-voxel resolution map $\mathcal{R}_\mathrm{map}$; mesh TSDF $d_{\mathcal{M}}$; flow--freeze prior $\{\mathcal{B}_\mathrm{flow}^\mathrm{hier},\mathcal{B}_\mathrm{freeze}^\mathrm{hier}\}$
\hfill\Comment{App.~\ref{app:sdf}}
\Ensure Accepted primitives $\mathcal{Q}$, claimed volume $\Omega_{\mathrm{claimed}}$

\Statex \textbf{// Initialization}
\State $\mathrm{flow\_area}\leftarrow\mathcal{B}_\mathrm{flow}^\mathrm{hier}$;\;
       $\mathrm{freeze\_area}\leftarrow\mathcal{B}_\mathrm{freeze}^\mathrm{hier}$
\State $\mathrm{occupied}\leftarrow\mathrm{flow\_area}\cup\mathrm{freeze\_area}$;\;
       $\mathrm{ext\_void}\leftarrow\neg\,\mathrm{fillholes}(\mathrm{occupied})$
\State $\Omega_\mathrm{freeze}\leftarrow\mathrm{freeze\_area}\cup\mathrm{ext\_void}$
\hfill\Comment{Freeze zone}
\State $\Omega_\mathrm{flow}\leftarrow\mathrm{dilate}(\neg\,\mathrm{ext\_void},\,2\Delta^{r_\mathrm{fit}})$;\;
       $\mathcal{Q},\,\Omega_\mathrm{claimed}\leftarrow\emptyset$
\hfill\Comment{Flow zone}

\Statex \textbf{// Coarse-to-fine fitting stages}
\For{$r\in\{16,32,64\}$ \textbf{from coarse to fine}}
    \State $r_\mathrm{fit}\leftarrow\min(r,32)$;\;
           $\Omega_\mathrm{valid}\leftarrow\{\mathcal{R}_\mathrm{map}\!\le\!r\}\cap\neg\Omega_\mathrm{freeze}$
    \State $\Omega_\mathrm{spec}\leftarrow\{\mathcal{R}_\mathrm{map}\!>\!r\}$
    \hfill\Comment{reserved for finer stages}
    \While{$\mathrm{coverage}(\Omega_\mathrm{valid})<0.95$}
        \State $\mathbf{s}\leftarrow\textsc{BodyOrSurfaceSeed}(\Omega_\mathrm{valid}\setminus\Omega_\mathrm{claimed})$;\;\textbf{break} if $\mathbf{s}{=}\varnothing$
        \State $\mathrm{ROI}\leftarrow\mathrm{ball}(\mathbf{s},\rho(r_\mathrm{fit}))\cap\Omega_\mathrm{valid}\setminus(\Omega_\mathrm{claimed}\cup\Omega_\mathrm{freeze})$
        \State $\mathcal{P}_{\mathbf{s}}\leftarrow\{\mathbf{p}\in\mathrm{ROI}:d_{\mathcal{M}}(\mathbf{p})<1.5\Delta\}$
        \hfill\Comment{interior + thin outer band}
        \State $(\boldsymbol{\theta}_0,\mathrm{lb},\mathrm{ub})\leftarrow\textsc{InitSQ}(\mathbf{s},r_\mathrm{fit})$
        \hfill\Comment{init radius capped by $\mathrm{EDT}_{\Omega_\mathrm{flow}}(\mathbf{s})$}
        \For{$\eta\in\{1.0,0.5,0.25\}$ \textbf{(retry loop)}}
            \State $\boldsymbol{\theta}^{*}\leftarrow\textsc{FitSDF}(\mathcal{P}_\mathbf{s},d_{\mathcal{M}},\boldsymbol{\theta}_0;\,a_\mathrm{scale}{=}\eta)$
            \hfill\Comment{Alg.~\ref{alg:fit_sdf}}
            \If{$\partial\Omega(\boldsymbol{\theta}^{*})\subseteq\Omega_\mathrm{flow}$} \textbf{break} \EndIf
        \EndFor
        \State \textbf{if} loop exhausted \textbf{then} reject seed; \textbf{continue}
        \State $M_\mathrm{fit}\leftarrow\{\mathbf{v}:\tilde d_{\boldsymbol{\theta}^{*}}(\mathbf{v})\le 2\Delta\}$;\;
               $\Delta\Omega\leftarrow M_\mathrm{fit}\setminus(\Omega_\mathrm{freeze}\cup\Omega_\mathrm{claimed}\cup\Omega_\mathrm{spec})$
        \If{$|\Delta\Omega|\ge V_\mathrm{min}$}
            \State $\mathcal{Q}\leftarrow\mathcal{Q}\cup\{\boldsymbol{\theta}^{*}\}$;\;
                   $\Omega_\mathrm{claimed}\leftarrow\Omega_\mathrm{claimed}\cup(M_\mathrm{fit}\setminus\Omega_\mathrm{freeze})$
            \hfill\Comment{hard clipping}
        \EndIf
    \EndWhile
\EndFor
\State \Return $\mathcal{Q},\,\Omega_\mathrm{claimed}$
\hfill\Comment{small residuals merged into nearest primitive}
\end{algorithmic}
\end{algorithm}

\begin{algorithm}[!t]
\caption{\textsc{FitSDF}: Region-Weighted TSDF Fitting Kernel}
\label{alg:fit_sdf}
\begin{algorithmic}[1]
\Require Sample set $\mathcal{P}$ with TSDF values $\{d_{\mathcal{M}}(\mathbf{p})\}$; initial parameters $\boldsymbol{\theta}_0$ with bounds $[\mathrm{lb},\mathrm{ub}]$; truncation $\tau$; semi-axis trust-region scale $a_\mathrm{scale}$
\Ensure Optimized parameters $\boldsymbol{\theta}^{*}$

\State $\mathcal{P}_\mathrm{a}\leftarrow\{\mathbf{p}\in\mathcal{P}:|\tilde d_{\boldsymbol{\theta}_0}(\mathbf{p})|<3\tau\}$
\hfill\Comment{active set}
\For{$\mathbf{p}\in\mathcal{P}_\mathrm{a}$}
    \If{$d_{\mathcal{M}}(\mathbf{p})>0$}
        \State $w(\mathbf{p})\leftarrow 1+\mu_\mathrm{ext}\,\max(0,-\tilde d_{\boldsymbol{\theta}_0}(\mathbf{p}))/\tau$
        \hfill\Comment{exterior intrusion penalty}
    \ElsIf{$|d_{\mathcal{M}}(\mathbf{p})|<0.3\tau$}
        \State $w(\mathbf{p})\leftarrow 1$
        \hfill\Comment{near-surface}
    \Else
        \State $w(\mathbf{p})\leftarrow w_\mathrm{Gauss}(\mathbf{p})$
        \hfill\Comment{interior Gaussian inlier weight}
    \EndIf
\EndFor
\State Define residual $r_{\boldsymbol{\theta}}(\mathbf{p})=\tilde d_{\boldsymbol{\theta}}(\mathbf{p})-d_{\mathcal{M}}(\mathbf{p})$, clipped to $\pm 0.3\tau$ on near-surface points and $\pm\tau$ otherwise
\hfill\Comment{two-stage truncation}
\State $\boldsymbol{\theta}^{*}\leftarrow\arg\min_{\boldsymbol{\theta}\in[\mathrm{lb},\mathrm{ub}]}\sum_{\mathbf{p}\in\mathcal{P}_\mathrm{a}} w(\mathbf{p})\,r_{\boldsymbol{\theta}}(\mathbf{p})^{2}$
\hfill\Comment{TRF least-squares; $x_\mathrm{scale}[a_x,a_y,a_z]{=}a_\mathrm{scale}$}
\State $c\leftarrow\bigl(\sum_{\mathbf{p}}w(\mathbf{p})\,r_{\boldsymbol{\theta}^{*}}(\mathbf{p})^{2}\bigr)/|\mathcal{P}_\mathrm{a}|$
\State \Return $\boldsymbol{\theta}^{*},\,c$
\end{algorithmic}
\end{algorithm}

\section{More Visualization Cases}
\label{app:visual}
\begin{figure*}[ht]
    \centering
    \includegraphics[width=0.95\textwidth]{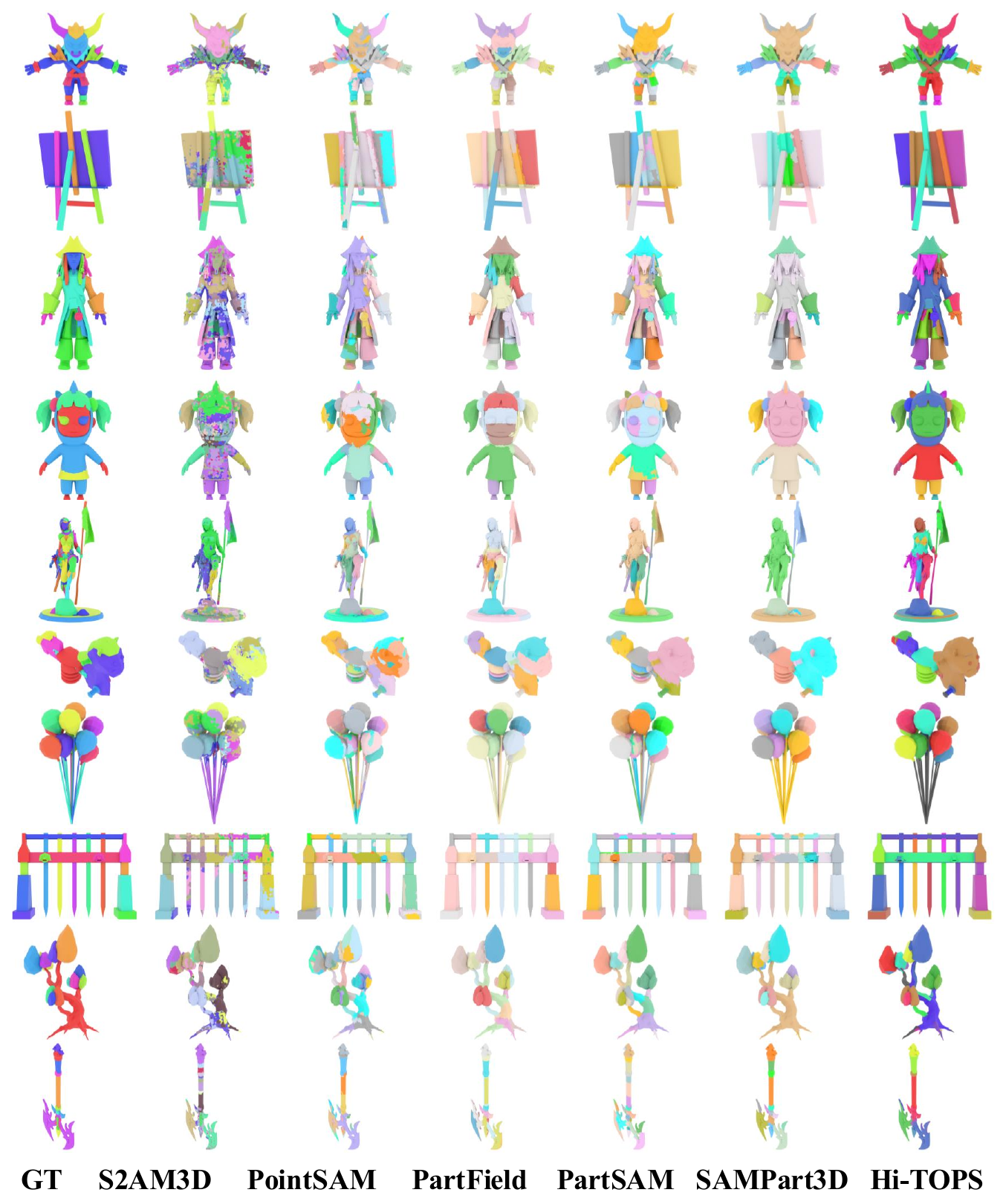}
    \caption{More qualitative comparison case results. Our method demonstrates competitive segmentation performance with topologically faithful structure.}
    \label{fig:app_visual_compare}
\end{figure*}

Fig.~\ref{fig:app_visual_compare} shows the seams produced by Hi-TOPS remain clean and well aligned with the underlying geometric ridges across diverse shapes.

\end{document}